\documentclass[twoside,twocolumn,9pt]{article}
\usepackage{extsizes}
\usepackage[super,sort&compress,comma]{natbib} 
\usepackage[version=3]{mhchem}
\usepackage[left=1.5cm, right=1.5cm, top=1.785cm, bottom=2.0cm]{geometry}
\usepackage{balance}
\usepackage{mathptmx}
\usepackage{sectsty}
\usepackage{graphicx} 
\usepackage{lastpage}
\usepackage[format=plain,justification=justified,singlelinecheck=false,font={stretch=1.125,small,sf},labelfont=bf,labelsep=space]{caption}
\usepackage{float}
\usepackage{fancyhdr}
\usepackage{fnpos}
\usepackage[english]{babel}
\addto{\captionsenglish}{%
  \renewcommand{\refname}{References}
}
\usepackage{array}
\usepackage{droidsans}
\usepackage{charter}
\usepackage[T1]{fontenc}
\usepackage[usenames,dvipsnames]{xcolor}
\usepackage{setspace}
\usepackage[compact]{titlesec}
\usepackage{hyperref}
\usepackage{multirow}%
\usepackage{amsmath,amssymb,amsfonts}%
\usepackage{amsthm}%
\usepackage{mathrsfs}%
\usepackage[title]{appendix}%
\usepackage{xcolor}%
\usepackage{textcomp}%
\usepackage{manyfoot}%
\usepackage{booktabs}\let\cline\cmidrule%
\usepackage{algorithm}%
\usepackage{algorithmicx}%
\usepackage{algpseudocode}%
\usepackage{listings}%
\usepackage{dcolumn}
\usepackage{bm}
\usepackage{hyperref}
\usepackage[capitalise]{cleveref}
\usepackage{makecell}
\usepackage{xcolor}   
\usepackage{wasysym}
\usepackage{tikzsymbols}
\usepackage{multirow}
\usepackage{lipsum}
\usepackage{subcaption}
\usepackage{tabularx}
\usepackage{multibib}
\usepackage{algpseudocode}
\newcites{S}{Supplementary References}
\usepackage{pifont}
\newcommand{\cmark}{\ding{51}}%
\newcommand{\xmark}{\ding{55}}%

\algdef{SE}[VARIABLES]{Variables}{EndVariables}
   {\algorithmicvariables}
   {\algorithmicend\ \algorithmicvariables}
\algnewcommand{\algorithmicvariables}{\textbf{parameter description}}

\definecolor{cream}{RGB}{222,217,201}
\begin{document}

\makeFNbottom
\makeatletter
\renewcommand\LARGE{\@setfontsize\LARGE{20pt}{17}}
\renewcommand\Large{\@setfontsize\Large{12pt}{14}}
\renewcommand\large{\@setfontsize\large{10pt}{12}}
\renewcommand\footnotesize{\@setfontsize\footnotesize{7pt}{10}}
\makeatother

\renewcommand{\thefootnote}{\fnsymbol{footnote}}
\renewcommand\footnoterule{\vspace*{1pt}%
\color{cream}\hrule width 3.5in height 0.4pt \color{black}\vspace*{5pt}} 
\setcounter{secnumdepth}{5}

\makeatletter 
\renewcommand\@biblabel[1]{#1}            
\renewcommand\@makefntext[1]%
{\noindent\makebox[0pt][r]{\@thefnmark\,}#1}
\makeatother 
\renewcommand{\figurename}{\small{Fig.}~}
\sectionfont{\sffamily\Large}
\subsectionfont{\normalsize}
\subsubsectionfont{\bf}
\setstretch{1.125} 
\setlength{\skip\footins}{0.8cm}
\setlength{\footnotesep}{0.25cm}
\setlength{\jot}{10pt}
\titlespacing*{\section}{0pt}{4pt}{4pt}
\titlespacing*{\subsection}{0pt}{15pt}{1pt}

\makeatletter 
\newlength{\figrulesep} 
\setlength{\figrulesep}{0.5\textfloatsep} 

\newcommand{\topfigrule}{\vspace*{-1pt}%
\noindent{\color{cream}\rule[-\figrulesep]{\columnwidth}{1.5pt}} }

\newcommand{\botfigrule}{\vspace*{-2pt}%
\noindent{\color{cream}\rule[\figrulesep]{\columnwidth}{1.5pt}} }

\newcommand{\dblfigrule}{\vspace*{-1pt}%
\noindent{\color{cream}\rule[-\figrulesep]{\textwidth}{1.5pt}} }

\makeatother

\twocolumn[
  \begin{@twocolumnfalse}

 \begin{tabular}{m{1.6cm} p{13.8cm} }

  & \noindent\LARGE{\textbf{Guided Multi-objective Generative AI to Enhance Structure-based Drug Design}} \\
\vspace{0.3cm} & \vspace{0.3cm} \\
& \noindent\large{Amit Kadan $^{\ast}$\textit{$^{a}$}, Kevin Ryczko $^{\ast}$\textit{$^{a}$}, Erika Lloyd\textit{$^{a}$}, Adrian Roitberg\textit{$^{b}$}, Takeshi Yamazaki$^{\ast}$\textit{$^{a}$}}\\
\vspace{0.5cm} & \vspace{0.5cm} \\
\vspace{0.5cm} & \vspace{0.5cm} \\
& \noindent\normalsize{Generative AI has the potential to revolutionize drug discovery. Yet, despite recent advances in deep learning, existing models cannot generate molecules that satisfy all desired physicochemical properties. Herein, we describe IDOLpro, a novel generative chemistry AI combining diffusion with multi-objective optimization for structure-based drug design. Differentiable scoring functions guide the latent variables of the diffusion model to explore uncharted chemical space and generate novel ligands \textit{in silico}, optimizing a plurality of target physicochemical properties. 
We demonstrate our platform's effectiveness by generating ligands with optimized binding affinity and synthetic accessibility on two benchmark sets. 
IDOLpro produces ligands with binding affinities over 10\%-20\% higher than the next best state-of-the-art method on each test set, producing more drug-like molecules with generally better synthetic accessibility scores than other methods. We do a head-to-head comparison of IDOLpro against an exhaustive virtual screen of a large database of drug-like molecules. We show that IDOLpro can generate molecules for a range of important disease-related targets with better binding affinity and synthetic accessibility than any molecule found in the virtual screen while being over $100\times$ faster and less expensive to run.
On a test set of experimental complexes, IDOLpro is the first to produce molecules with better binding affinities than the experimentally observed ligands. IDOLpro can accommodate other scoring functions (e.g. ADME-Tox) to accelerate hit-finding, hit-to-lead, and lead optimization for drug discovery.}

\end{tabular}

 \end{@twocolumnfalse} \vspace{0.6cm}
  ]

\renewcommand*\rmdefault{bch}\normalfont\upshape
\rmfamily
\vspace{-1cm}

\footnotetext{\textit{$^{a}$~SandboxAQ, Palo Alto, CA, United States}}

\footnotetext{\textit{$^{b}$~{Department of Chemistry, University of Florida, PO Box 117200, Gainesville, 2611-7200, Florida, United States. }}}
\footnotetext{\textit{$^{*}$Corresponding authors: amit.kadan@sandboxaq.com, kevin.ryczko@sandboxaq.com, takeshi.yamazaki@sandboxaq.com}}

\section{Introduction}
The central goal of structure-based drug design (SBDD) is to design ligands with high binding affinities to a target protein pocket given the 3-dimensional information\cite{anderson2003process}. SBDD is inherently an inverse design problem, where the desired properties (high binding affinity to a target protein, synthesizability, etc.) are known, but the design of a molecule with the desired properties is non-trivial. Inverse design problems are prevalent in the materials \cite{zunger2018inverse, ryczko2020inverse, cornet2023inverse}, chemical \cite{sanchez2017optimizing, gebauer2022inverse, sridharan2022modern}, and life sciences \cite{lee2023exploring, zaman2023stride, weiss2023guided}. They have two fundamental steps: The sampling of a chemical space, and the evaluation of compounds' abilities to satisfy the set of desired properties. 
The sampling of chemical space can be done in various ways. For drug discovery, this is typically done by evaluating each entry of a large database of drug-like molecules such as ZINC \cite{irwin2005zinc}, Enamine \cite{shivanyuk2007enamine}, or GDB~\cite{ruddigkeit2012enumeration}, collecting the results and ranking them to yield a shortlist of compounds to be screened in a laboratory. This approach is time-consuming and is restricted to searching through a fraction of drug-like chemical space -- despite some of these databases contain hundreds of billions of molecules~\cite{ruddigkeit2012enumeration}, drug-like chemical space is estimated to number between $10^{20}-10^{60}$ molecules~\cite{polishchuk2013estimation}. 

\begin{figure*}[ht]
    \centering
    \includegraphics[width=\linewidth]{figs/1.pdf}
    \caption{Visual overview of IDOLpro. A random latent vector, $\mathbf{z}_T^{\ell}$, is sampled for each ligand in the batch conditioned on the pocket coordinates, $\mathbf{z}^p$. Reverse diffusion is run from time $T$ to the optimization horizon, time $t_{hz}$. The rest of the diffusion process is completed, and the ligands are scored by evaluating a set of differentiable scores defining target physicochemical properties. The gradient with respect to each latent vector at the optimization horizon, $\partial S_i / \partial \mathbf{z}_{t_{hz}}^{\ell}$, is used to take a gradient step in the latent space. This process is iterated until a maximum number of steps have been reached, or a valid ligand cannot be generated with the current latent vector.}
    \label{fig:overview}
\end{figure*}

Deep-learning (DL) based generative models for SBDD can replace virtual screening by directly predicting high-affinity ligands for a given protein pocket\cite{luo20213d, peng2022pocket2mol, graphbp, targetdiff, schneuing2022structure, li2021structure}. Initially, these models were predominantly autoregressive. Ref.~\cite{li2021structure} develop a sampling scheme based on Monte-Carlo tree search to condition an autoregressive generative model to generate molecules directly into the protein pocket. Ref.~\cite{peng2022pocket2mol} train an equivariant graph neural network called Pocket2Mol to predict novel molecules in a protein pocket by sequentially predicting atoms of the ligand given the current atomic context. Recently, several diffusion-based generative models have been proposed~\cite{targetdiff,schneuing2022structure}. Diffusion models learn to sample molecules by learning to reverse a diffusive noising process~\cite{ho2020denoising}. 
Unlike autoregressive techniques, these models allow for consideration of global relationships between atoms in the ligand throughout generation. Ref.~\cite{targetdiff} introduces an equivariant diffusion model called TargetDiff which generates ligands from scratch directly in the protein pocket. They also showed that with proper parameterization, their model can be used as an unsupervised binding affinity predictor, allowing them to filter promising candidates during generation. Similarly, Ref.~\cite{schneuing2022structure} introduces two equivariant diffusion models for SBDD. On top of de-novo generation, these models are able to perform scaffold-hopping and fragment merging, both of which are facilitated by an inpainting-inspired molecular generation scheme~\cite{lugmayr2022repaint}.

DL-based generative models are more efficient than database iteration and can produce molecules that don't currently exist in commercial databases, expanding the chemical space available during the drug-discovery process~ \cite{li2021structure}. However, in some cases they have been shown to produce ligands that do not exhibit desired physicochemical properties, producing compounds with unphysical structures \cite{buttenschoen2024posebusters,yu2023deep}, or producing compounds that exhibit poor synthesizability~\cite{gao2020synthesizability}. Just like in database generation, to operate in an inverse design pipeline, molecules generated by these models must be filtered and ranked with physicochemical descriptors, and there is no guarantee of finding a molecule satisfying a set of desired properties.

Alternatively, guidance from physicochemical scores can be directly integrated into the generation process~\cite{gomez2018automatic, gebauer2022inverse, lee2023exploring}. 
Ref. \cite{li2021structure} pair a policy network with Monte Carlo Tree Search to design ligands with optimized Vina scores. They apply their methodology to design an inhibitor for the main protease of SARS-CoV-2, producing novel ligands with high binding affinity. Ref.~\cite{schneuing2022structure} pairs a diffusion model with an evolutionary algorithm to optimize the properties of generated ligands. Ref.~\cite{dollar2023efficient} uses property predictors to guide a generative model composed of an equivariant autoencoder and transformer decoder in generating molecules with good target properties. 
These works do not make use of gradient information in their optimization schemes. Gradient information forms the basis of many modern optimization algorithms~\cite{bubeck2015convex}, particularly in DL applications, allowing for optimal scaling with the degrees of freedom of the  optimization~\cite{duchi2015optimal}. Ref.~\cite{dhariwal2021diffusion} introduces classifier guidance, a technique that allows diffusion models to generate samples conditioned on the gradients of classifiers throughout the reverse diffusion process. Similarly, one can use gradients from regressors trained to quantify properties of interest. Regressor guidance has been implemented in recent works on generative molecular design~\cite{ziv2024molsnapper, weiss2023guided, lee2023exploring}. Ref~\cite{ziv2024molsnapper} use regressor guidance to re-purpose a generative model trained to generate gas-phase molecules for structure-based drug design.  Ref.~\cite{weiss2023guided} use both regressor and classifier guidance to generate optimized polyiclic aromatic systems -- molecules important for the design of organic semiconductors. Ref~\cite{lee2023exploring} use regressor guidance to design drug-like molecules with optimized target properties including protein-ligand interaction, drug-likeness, and synthesizability. A drawback of regressor guidance is that it requires training ad-hoc property predictors on intermediate states in the reverse diffusion of a particular generative model, and cannot be used with an abundance of state-of-the-art physicochemical scores and models designed to make predictions on three-dimensional molecules~\cite{trott2010autodock,thakkar2021retrosynthetic,corso2022diffdock}.

In this work, we present IDOLpro (\textbf{I}nverse \textbf{D}esign of \textbf{O}ptimal \textbf{L}igands for \textbf{Pro}tein pockets), a generative workflow that actively guides a diffusion model to generate an optimized set of ligands for a target protein pocket. Specifically,
we modify the latent variables of a diffusion model at a carefully chosen time step in the reverse diffusion process to optimize one or more objectives of interest simultaneously. We achieve this by backpropagating gradients obtained by evaluating physicochemical scores directly on the generated molecular structures.
This avoids
the need for developing ad-hoc regressors capable of making predictions on intermediate noisy representations, and allows us to directly leverage common scores for assessing ligand quality. In this report, we include binding affinity and synthetic accessibility, but our framework is highly modular and can incorporate alternative generators and additional scores.
All metrics are written in PyTorch \cite{paszke2019pytorch} and are fully differentiable with respect to the latent variables within the diffusion model, allowing for the use of gradient-based optimization strategies to design optimal ligands.

\begin{table*}[ht!]
\resizebox{\linewidth}{!}{%
\centering
\begin{tabular}{c|c|c|c|c|c|c|c|c|c}
    Dataset & Method & Vina [kcal/mol] & Vina$_{10\%}$ [kcal/mol] &  SA & SA$_{10\%}$ & Synth [\%] & QED & Diversity & PB-valid [\%]\\ \Xhline{2\arrayrulewidth}
    \multirow{2}{*}{CrossDocked} 
    & DiffSBDD-cond~\cite{schneuing2022structure} & $-5.37 \pm 1.93$ & $-7.70 \pm 2.25$  & $4.30 \pm 0.50$ & $2.56 \pm 0.63$ & $23.5 \pm 15.5$ & $0.56 \pm 0.05$ & $0.78 \pm 0.07$ & $55.9 \pm 11.1$ \\
    & IDOLpro & $-6.47 \pm 2.10$ & $-8.69 \pm 2.45$ & $3.41 \pm 0.70$ & $1.73 \pm 0.51$ & $51.2 \pm 22.7$ & $0.63 \pm 0.06$ & $0.79 \pm 0.07$ & $64.3 \pm 8.3$ \\
    \hline
    
    \multirow{2}{*}{MOAD} 
    & DiffSBDD-cond~\cite{schneuing2022structure} & $-5.38 \pm 2.55$ & $-7.66 \pm 2.37$ & $4.18 \pm 0.50$ & $2.50 \pm 0.53$ & $26.5 \pm 17.8$ & $0.58 \pm 0.07$ & $0.77 \pm 0.07$ & $53.6 \pm 15.7$ \\
    & IDOLpro & $-6.77 \pm 2.24$ & $-8.68 \pm 2.56$ & $3.32 \pm 0.66$ & $1.79 \pm 0.46$ & $56.9 \pm 22.6$ & $0.63 \pm 0.08$ & $0.77 \pm 0.07$ & $60.4 \pm 16.3$
\end{tabular}}
\caption{Performance of IDOLpro when used to optimize torchvina and torchSA relative to the baseline model, DiffSBDD-cond on the CrossDocked and Binding MOAD test sets. The average Vina score, top-10\% Vina score, SA score, top-10\% SA score, percent of synthesizable molecules, QED, diversity, and PoseBusters validity are reported.}
\label{tab:de_novo_improve}
\end{table*}

\begin{table*}[ht!]
\centering
\begin{tabular}{c|c|c|c|c|c|c|c}
    Method & SA & SA$_{10\%}$ & Synth [\%] & QED & Diversity & PB-valid [\%] & Time [s/ligand] \\ 
    \Xhline{2\arrayrulewidth}
    Regressor guidance & $3.53 \pm 0.37$ & $1.38 \pm 0.18$ & $50.6 \pm 10.5$ & $0.37 \pm 0.01$  & $0.90 \pm 0.02$ & $7.6 \pm 3.6$ & $8.11 \pm 1.64$ \\
    IDOLpro & $2.79 \pm 0.55$ & $1.53 \pm 0.40$ & $71.51 \pm 18.6$ & $0.62 \pm 0.06$ & $0.81 \pm 0.06$ & $43.7 \pm 16.2$ & $82.33 \pm 52.51$  \\
\end{tabular}
\caption{Performance of IDOLpro compared to DiffSBDD regressor guidance on the CrossDocked test set. The average SA, QED, diversity, PoseBusters validity, and average time to generate an individual molecule are reported.}
\label{tab:regressor_guide}
\end{table*}

\section{Results}

\subsection{Workflow}

IDOLpro takes the protein pocket information as input and iteratively modifies the predictions of a generator that generates molecules directly into the pocket to produce optimal ligands according to a set of differentiable scores defining target properties. IDOLpro accomplishes this by modifying the latent vectors of the generative model with gradients from the property predictors. As shown in \cref{fig:overview}, IDOLpro freezes the latent vectors at a specified time step in reverse diffusion -- the \textit{optimization horizon} denoted by $t_{hz}$. IDOLpro then unwinds the rest of the reverse diffusion process, using gradients from differentiable property predictors to update the frozen latent vectors. This procedure is repeated iteratively, resulting in an improved sample with respect to the scores at the end of the optimization. Once a set of optimized ligands is produced, their binding poses are refined by local structural optimization within the protein pocket. Structural refinement interfaces with the same differentiable scores used for latent optimization and iteratively modifies a ligand's coordinates with gradients from the property predictors.

In this report, our method uses DiffSBDD~\cite{schneuing2022structure} as the baseline generative method for predicting ligands within a protein pocket. We use a specific variant of the model, DiffSBDD-Cond, which we found was able to generate ligands within the protein pocket without clashes more reliably than the alternative, DiffSBDD-inpaint. We assess the ability of our framework to discover novel ligands with improved binding affinity and synthetic accessibility. To evaluate and be able to gather gradient information for binding affinity, we have developed a torch-based implementation of the Vina score~\cite{trott2010autodock}, which we refer to as torchvina. To evaluate synthetic accessibility, we have trained an equivariant neural network model to predict the synthetic accessibility (SA) score first proposed in Ref.~\cite{ertl2009estimation}, which we refer to as torchSA. During structural refinement, we also make use of the ANI2x~\cite{devereux2020extending} neural network potential, which we use to optimize intramolecular forces, allowing IDOLpro to produce physically valid molecules.

\subsection{Datasets}
We assess the performance of IDOLpro on three different test sets -- a subset of CrossDocked~\cite{francoeur2020three}, a subset of the Binding MOAD (Mother of all Databases)~\cite{hu2005binding}, and on a test set consisting of disease-related proteins first proposed in Ref.~\cite{fu2022reinforced}, which we refer to as the RGA test set. 

All three databases contain protein pocket-ligand pairs. The CrossDocked dataset contains 100 pocket-ligand pairs which were derived via re-docking ligands to non-cognate receptors with smina~\cite{koes2013lessons}. This test set was used to validate the performance of tools in several other papers~\cite{luo20213d,peng2022pocket2mol}, including DiffSBDD~\cite{schneuing2022structure}. The Binding MOAD contains 130 high-resolution ($<$2.5 Å) experimentally derived pocket-ligand pairs extracted from the Protein Data Bank (PDB). This test set was also used to assess the performance of DiffSBDD. The RGA test set consists of 10 disease-related protein targets including G-protein coupling receptors (GPCRs) and kinases from the DUD-E dataset~\cite{mysinger2012directory}, as well as the SARS-CoV-2 main protease~\cite{zhang2021potent}. This is also an experimentally derived test set and was used to assess the performance of several non-deep learning-based methods in Ref~\cite{fu2022reinforced}.

\subsection{Validation of Latent Vector Optimization}\label{subsec:latent-opt} 

To assess the ability of IDOLpro to augment the performance of the baseline model via the optimization of latent vectors, we run IDOLpro while optimizing torchvina and torchSA,  and
analyze its capability to improve the Vina and SA scores relative
to DiffSBDD-Cond. For each of the protein pockets in the CrossDocked and Binding MOAD test sets, we generate 100 optimized ligands using IDOLpro. Generated ligands do not undergo structural refinement to isolate the effect of optimizing latent vectors.

We report a number metrics averaged across the protein pockets in each test set. Each metric is recorded before and after optimization with IDOLpro. Reported metrics include the Vina and top-10\% Vina scores of generated ligands, and three metrics using the SA score -- the average SA, top-10\% SA and the percentage of synthesizable molecules generated. In this work, we define a ligand as synthesizable if it achieves an SA score of less than 3.5. Although the inventors of the SA score suggest 6 as the cutoff for synthesizability, a number of papers have found SA scores between 3.5 and 6 to be ambiguous~\cite{gao2020synthesizability,thakkar2021retrosynthetic}. A cutoff of 3.5 was used to determine synthesizability in Ref.~\cite{yu2022organic}. We make note of the average QED (quantitative estimate of drug-likeness)~\cite{bickerton2012quantifying}, a metric combining several desirable molecular properties for screening drug-like molecule, and pocket diversity -- the average pairwise dissimilarity between all generated molecules for a given protein pocket. Dissimilarity is measured as $1 - \text{Tanimoto similarity}$. Lastly, we use PoseBusters~\cite{buttenschoen2024posebusters} to assess the validity of generated molecules. PoseBusters assesses the validity of generated molecules by running a series of checks using RDKit~\cite{rdkit}, ensuring stereochemistry and intra- and inter-molecular constraints such as appropriate bond lengths, planarity of aromatic rings, and no overlap with the protein pocket are satisfied.

The results are shown in \cref{tab:de_novo_improve}. We find that IDOLpro generates ligands with better Vina and SA scores than DiffSBDD-cond, yielding molecules with $\approx 20 \%$ better Vina scores and $\approx 21 \%$ better SA scores on the CrossDocked test set, and $\approx 26 \%$ better Vina scores and $\approx 21 \%$ better SA scores on the Binding MOAD test set. Furthermore, IDOLpro yields more than double the amount of synthesizable ligands compared with DiffSBDD-cond for each dataset (51.2 \% vs 23.5 \% and 56.9 \% vs 22.6 \%). Lastly, despite not optimizing for these directly, we find that IDOLpro generates molecules with slightly higher QED and PoseBusters validity (PB-valid in \cref{tab:de_novo_improve}) than DiffSBDD-Cond, achieving higher values on both the CrossDocked and Binding MOAD test sets. Despite IDOLpro performing an optimization in latent space, it does not decrease the diversity of generated molecules, achieving equal average pocket diversity to DiffSBDD-Cond on each test set.

We visualize a single example from each benchmark test set in \cref{fig:de_novo_examples} . The figure shows an example of a molecule produced by DiffSBDD prior to latent vector optimization, and after latent vector optimization by IDOLpro. We also picture the corresponding reference molecule, showing that latent vector optimization allows IDOLpro to produce molecules with better metrics than the reference molecule. Overall, IDOLpro is able to co-optimize Vina and SA, producing molecules with significantly better binding affinities and synthetic accessibility when compared to the baseline model, DiffSBDD-cond.

\subsection{Comparison to Regressor Guidance}\label{subsec:reg-guide}

Next, we compare IDOLpro to regressor guidance -- a standard technique for steering the predictions of a diffusion model to optimize a set of scores~\cite{dhariwal2021diffusion,weiss2023guided,lee2023exploring,ziv2024molsnapper}. We run IDOLpro and DiffSBDD with regressor guidance to generate pocket-bound ligands while optimizing the SA score of generated moelcules. For each method, we generate 100 ligands for each protein pocket in the CrossDocked test set. We do not perform structural  refinement after generation for either method to isolate each technique's ability to generate optimized molecules. 

For each method, we report the SA score, top-10\% SA score, percentage of synthesizable molecules, QED, pocket diversity, and PoseBusters validity averaged across all protein pockets in the test set. We also report the average time required by each technique for generating a single ligand. Full results are showing in \cref{tab:regressor_guide}. We find that IDOLpro generates ligands with better average SA scores than DiffSBDD with regressor guidance, yielding molecules with $\approx 27 \%$ better SA scores (2.79 vs 3.53). IDOLpro also produces a significantly higher number of synthesizable molecules ($> 20\%$ more) than DiffSBDD with regressor guidance. On the other hand, DiffSBDD with regressor guidance achieves a better top-10\% SA score than IDOLpro. A major advantage of regressor guidance over IDOLpro is that it is $\approx 10\times$ faster for generating a single ligand. This is expected as regressor requires running reverse diffusion once, while IDOLpro unwinds the latter portion of reverse diffusion multiple times. IDOLpro yields molecules with significantly better QED and PoseBusters validity than DiffSBDD with regressor guidance, suggesting that adding gradients throughout reverse diffusion hinders the model's ability to adequately account for both intra- and inter-molecular constraints. This is corroborated by the fact that DiffSBDD without regressor guidance produces significantly more PoseBusters-valid molecules than DiffSBDD with regressor guidance (see \cref{tab:de_novo_improve} for vanilla DiffSBDD metrics).

\begin{figure*}[ht]
    \centering
    \begin{subfigure}{0.49\textwidth}
        \centering
        \includegraphics[width=\linewidth]{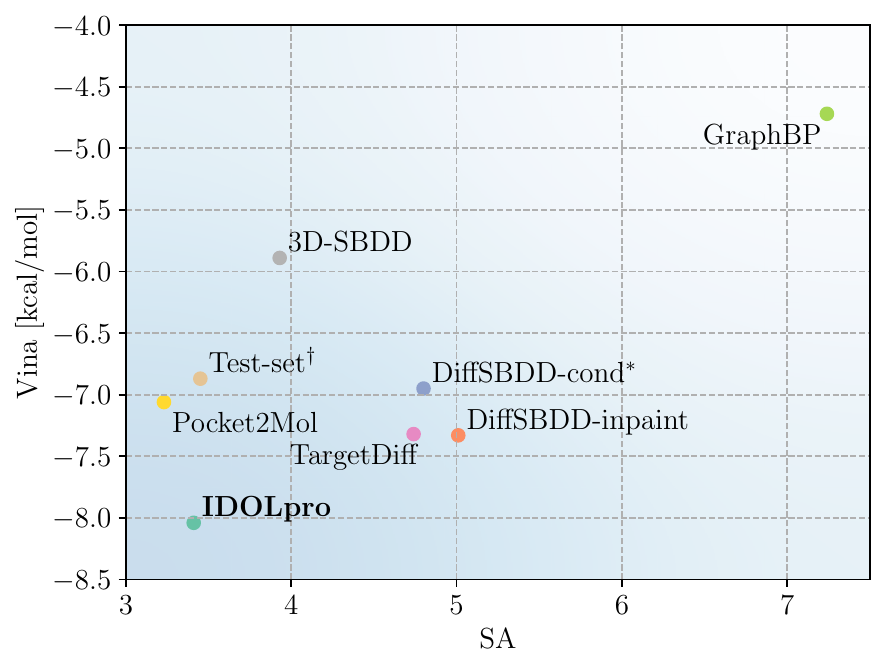}
    \end{subfigure}
    \begin{subfigure}{0.49\textwidth}
        \centering
        \includegraphics[width=\linewidth]{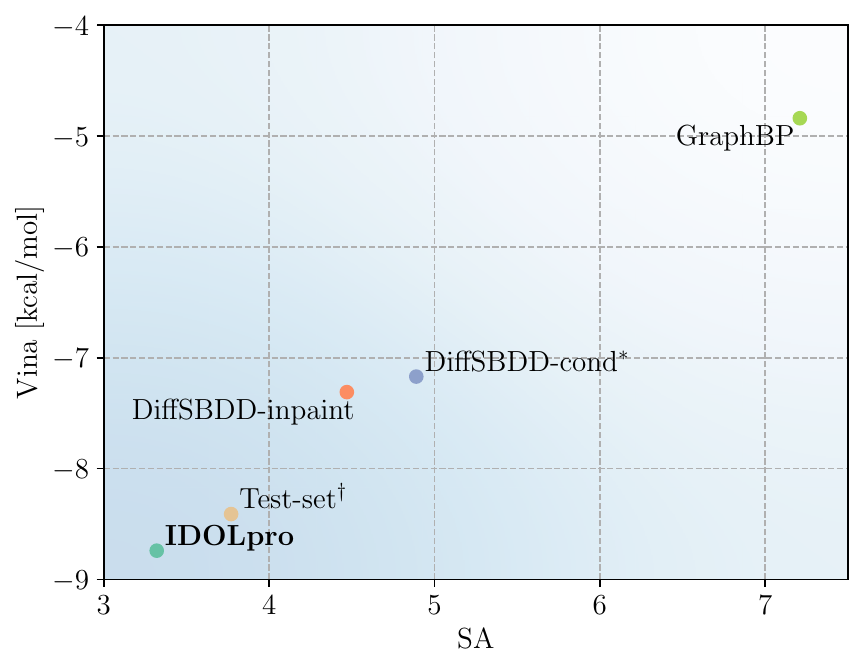}
    \end{subfigure}
    \caption{Performance of DL tools on two benchmark test sets. The scatter plot shows the average Vina and SA score for each method for targets in CrossDocked (left), and the Binding MOAD (right). IDOLpro is at the bottom left of each scatter plot, showing it can co-optimize Vina and SA for generated ligands. \\
    $\dagger$ Vina scores of reference ligands in the test set, redocked with QuickVina2.\\
    * Baseline method for IDOLpro.}
    \label{fig:benchmark_results}
\end{figure*}
\label{results}
\hspace*{-1.5cm}\begin{table*}[ht!]
\centering
\begin{tabular}{c|c|c|c|c|c|c|c}
    & Method & Vina [kcal/mol] & Vina$_{10\%}$ [kcal/mol] & SA & QED & Diversity & Time [s/ligand] \\ \Xhline{3\arrayrulewidth}
    \parbox[t]{2mm}{\multirow{9}{*}{\rotatebox[origin=c]{90}{CrossDocked}}} & Test Set & $-6.87 \pm 2.32$ & - & $3.45 \pm 1.26$ & $0.48 \pm 0.20$ & - & - \\ \cline{2-8}
    & 3D-SBDD~\cite{luo20213d} & $-5.89 \pm 1.91$ & $-7.29 \pm 2.34$ & $3.93 \pm 1.26$ & $0.50 \pm 0.17$ & $0.74 \pm 0.09$ & $328.13 \pm 245.43$ \\
    & Pocket2Mol~\cite{peng2022pocket2mol} & $-7.06 \pm 2.80$ & $-8.71 \pm 3.18$ & $\mathbf{3.23 \pm 1.08}$ & $0.57 \pm 0.16$ &  $0.74 \pm 0.15$ & $41.79 \pm 36.84$ \\
    & GraphBP~\cite{graphbp} & $-4.72 \pm 4.03$ & $-7.17 \pm 1.40$ & $7.24 \pm 0.81$ & $0.50 \pm 0.12$ & $\mathbf{0.84 \pm 0.01}$ & $\mathbf{0.17 \pm 0.02}$ \\
    & TargetDiff~\cite{targetdiff} & $-7.32 \pm 2.47$ & $-9.67 \pm 2.55$ & $4.74 \pm 1.17$ & $0.48 \pm 0.20$ & $0.72 \pm 0.09$ & $\sim 57.22$ \\
    & DiffSBDD-cond~\cite{schneuing2022structure} & $-6.95 \pm 2.06$ & $-9.12 \pm 2.16$ & $4.80 \pm 1.17$ & $0.47 \pm 0.21$ & $0.73 \pm 0.07$ & $2.27 \pm 0.86$  \\
    & DiffSBDD-inpaint~\cite{schneuing2022structure} & $-7.33 \pm 2.56$ & $-9.93 \pm 2.59$ & $5.01 \pm 1.08$ & $0.47 \pm 0.18$ &  $0.76 \pm 0.05$ & $2.67 \pm 1.22$ \\
    & IDOLpro & $\mathbf{-8.04 \pm 2.55}$ & $\mathbf{-10.96 \pm 3.02}$ & $3.41 \pm 0.70$ & $\mathbf{0.63 \pm 0.06}$ & $0.79 \pm 0.07 $ & $58.80 \pm 32.97$ \\
    \Xhline{2\arrayrulewidth}
    
    \parbox[t]{2mm}{\multirow{6}{*}{\rotatebox[origin=c]{90}{MOAD}}} & Test Set & $-8.41 \pm 2.03$ & - & $3.77 \pm 1.08$ & $0.52 \pm 0.17$ & - & - \\ \cline{2-8}
    & GraphBP~\cite{graphbp} & $-4.84 \pm 2.24$ & $-6.63 \pm 0.95$ & $7.21 \pm 0.81$ & $0.51 \pm 0.11$ & $\mathbf{0.83 \pm 0.01}$ & $\mathbf{0.23 \pm 0.03}$ \\
    & DiffSBDD-cond~\cite{schneuing2022structure} & $-7.17 \pm 1.89$ & $-9.18 \pm 2.23$ & $4.89 \pm 1.08$ & $0.44 \pm 0.20$ & $0.71 \pm 0.08$ & $5.61 \pm 1.42$  \\
    & DiffSBDD-inpaint~\cite{schneuing2022structure} & $-7.31 \pm 4.03$ & $-9.84 \pm 2.18$ & $4.47 \pm 1.08$ & $ 0.54 \pm 0.21$ & $0.74 \pm 0.05$ & $6.17 \pm 2.08$ \\
    & IDOLpro & $\mathbf{-8.74 \pm 2.59}$ & $\mathbf{-11.23 \pm 3.12}$ & $\mathbf{3.32 \pm 0.66}$ & $\mathbf{0.63 \pm 0.08}$ & $0.77 \pm 0.07$ & $82.30 \pm 45.07 $
\end{tabular}
\caption{Evaluation of DL tools on targets from the CrossDocked and Binding MOAD datasets. The average, along with the standard deviation of each metric across the protein pockets in each dataset is reported. The top performing model on each metric is bolded in the corresponding column. Numbers for other models are taken from Ref.~\cite{schneuing2022structure}. Time is based on running DiffSBDD-cond on our hardware, and adjusting the times reported in Ref.~\cite{schneuing2022structure} accordingly.
}
\label{tab:benchmark_results}
\end{table*}

\subsection{Comparison to Deep Learning}\label{subsec:compare-dl}

\begin{figure*}[ht!]
    \centering
    \includegraphics[width=\linewidth]{figs/3.pdf}
    \caption{Molecules produced by IDOLpro when optimizing torchvina and torchSA.  
    One example from each test set is shown -- protein 4aua from CrossDocked, and protein 3cjo from the Binding MOAD. Left column: reference molecules from the test sets. Middle column: initial ligand produced by DiffSBDD prior to latent vector optimization. Right column: molecule produced by IDOLpro after optimizing torchvina and torchSA. For each example, the molecule produced by DiffSBDD has worse Vina and SA scores than the reference molecule. After optimization with IDOLpro, both the Vina and SA scores of the generated molecule are better than the reference. Visualizations were created with PyMol~\cite{PyMOL}, and interactions were visualized with the protein-ligand interaction profiler (PLIP)~\cite{adasme2021plip}.} 
    \label{fig:de_novo_examples}
\end{figure*}

We compare IDOLpro to other deep learning tools in the literature on the CrossDocked and Binding MOAD test sets. As was done in Ref~\cite{schneuing2022structure}, we generate 100 optimized ligands using IDOLpro for each protein pocket. We report the Vina score, top-10\% Vina score, SA score, QED, pocket diversity, and time taken to generate a single ligand averaged across all protein pockets in each test. These metrics are evaluated for six other DL tools in the literature: 3D-SBDD~\cite{luo20213d}, Pocket2Mol~\cite{peng2022pocket2mol}, GraphBP~\cite{graphbp}, TargetDiff~\cite{targetdiff}, DiffSBDD-Cond, and DiffSBDD-inpaint~\cite{schneuing2022structure}. Results for other tools are taken from Ref.~\cite{schneuing2022structure}. 

In Ref.~\cite{schneuing2022structure}, the generated ligands of each deep learning model are re-docked to the target protein pocket with QuickVina2~\cite{alhossary2015fast}. In our workflow, we replace re-docking with structural refinement. In QuickVina2, molecular conformations are optimized using both global and local optimization, while structural refinement only performs local optimization on the ligand coordinates via L-BFGS. For more information on structural refinement, see Section~\cref{subsec:docking}.

Full results are shown in \cref{tab:benchmark_results}. The performance of the models on the two optimized metrics -- Vina and SA is visualized in \cref{fig:benchmark_results}. For CrossDocked, IDOLpro achieves greatly improved Vina scores relative to other DL tools, with a 0.71 kcal/mol improvement in average Vina score and  1.03 kcal/mol improvement in top-10\% Vina score compared to the next best tool in the literature, DiffSBDD-inpaint. Despite not optimizing QED directly, we find that IDOLpro produces ligands with the best QED out of all DL tools compared. IDOLpro ranks second for producing molecules with good SA scores, showing the ability of the platform to perform multi-objective optimization. Despite needing to run an entire optimization procedure for each ligand, IDOLpro is still computationally tractable, achieving run times competitive with two other tools -- TargetDiff and Pocket2Mol, while achieving a faster runtime than 3D-SBDD.

IDOLpro is almost always able to find ligands that improve upon both the Vina and SA scores relative to the reference ligand, failing to find such a ligand for only one target -- the protein pocket defined by ligand x2p bound to the protein with PDB ID 5mma. For this protein pocket, IDOLpro generated molecules with significantly better SA scores than the reference molecule, but did not generate a single molecule with a better Vina score. We hypothesize that re-weighting the objective in the optimization to more heavily favour the Vina score would allow IDOLpro to generate a ligand that improves upon both metrics.

For the Binding MOAD, the advantage of IDOLpro for generating molecules with high binding affinity is even more pronounced, with a 1.43 kcal/mol improvement in average Vina score, and 1.40 kcal/mol improvement in top-10\% Vina score compared to the next best method, DiffSBDD-inpaint. In particular, IDOLpro is the first ML tool to generate molecules with a better average Vina score than those of the reference molecules in the Binding MOAD test set. This is noteworthy, because unlike molecules in CrossDocked, molecules in the Binding MOAD were derived through experiment. Out of the four methods compared, IDOLpro also achieves the best SA, improving upon the next best method by 1.15, while also achieving the best QED despite not optimizing for it directly. The time to generate a single ligand for a protein pocket in the Binding MOAD test set is slower than for the CrossDocked test set, and reflects DiffSBDD's slowdown in proposing novel molecules for targets in this set. 

IDOLpro is able to find a molecule with a better Vina and SA score than the reference ligand for 126/130 targets ($\approx 97\%$ of the time). Similar to the one case in CrossDocked for which IDOLpro did not find a ligand with improved Vina and SA score relative to the reference ligand, for these 4 targets IDOLpro finds molecules with significantly better SA scores than the reference, but fails to find a molecule with a better Vina score. All four of these cases have reference molecules with SA scores of over 7, whereas IDOLpro does not produce a single molecule with an SA score over 6. We again hypothesize that this could be solved by re-weighting the optimization objective to more heavily favor Vina score.

Overall IDOLpro can effectively co-optimize multiple objectives, generating ligands with state-of-the-art binding affinity and synthetic accessibility on two test sets. Improving other metrics with IDOLpro is straightforward, simply requiring a differentiable score for evaluating the desired metric. Part of our future work will be to focus on including differentiable scoring functions for other desirable properties such as solubility, toxicity, etc.

\subsection{Comparison to Virtual Screening}\label{subsec:compare-vs}

\begin{figure*}[ht]
    \centering
    \begin{subfigure}{0.51\textwidth}
        \centering
        \includegraphics[width=\linewidth]{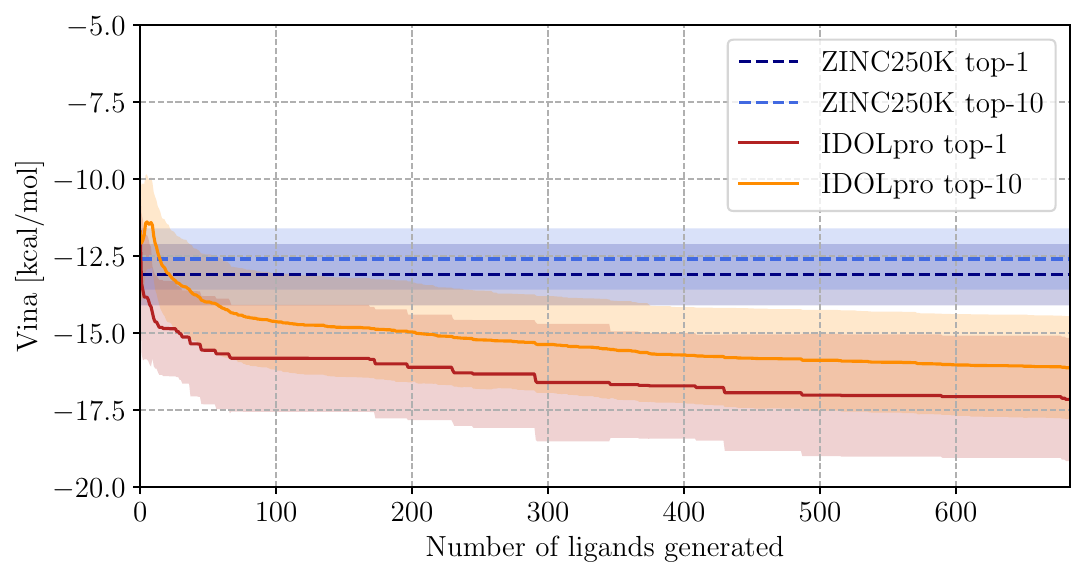}
    \end{subfigure}
    \begin{subfigure}{0.49\textwidth}
        \centering
        \includegraphics[width=\linewidth]{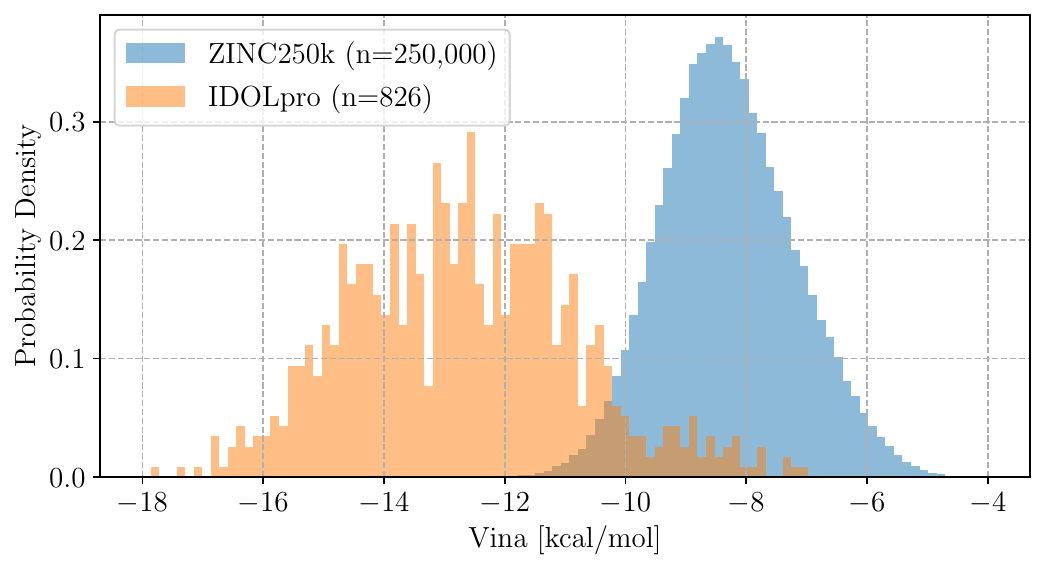}
    \end{subfigure}
    \begin{subfigure}{0.49\textwidth}
        \centering
        \includegraphics[width=\linewidth]{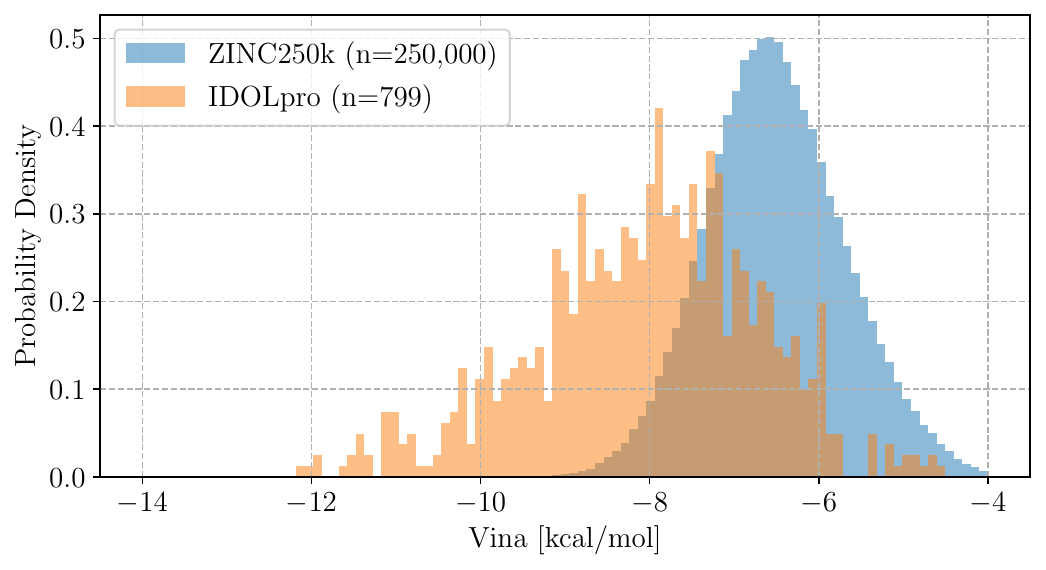}
    \end{subfigure}
    \caption{Comparison of IDOLpro to virtually screening ZINC250K. Top center: The average top-1 and top-10 Vina scores of IDOLpro generated molecules compared to screened molecules from ZINC250K across all 10 targets. Bottom left: The distribution of Vina scores for generated and screened ligands for EGFR, an important oncology target (PDB ID 2rgp). Bottom right: The distribution of Vina scores for generated and screened ligands for the SARS-Cov-2 main protease (PDB ID 7l11).}
    \label{fig:vs_vs_idolpro}
\end{figure*}

We compare the ability of IDOLpro to generate promising compounds when compared to an exahaustive virtual screen of a large library of commercially available drug-like compounds. To do so, we evaluate the performance of IDOLpro for finding ligands with high binding affinity and synthetic accessibility when compared with virtually screening the ZINC250K database -- a curated collection of 250,000 commercially available chemical compounds ~\cite{gomez2018automatic}. We evaluate each method on all 10 protein pockets in the RGA test set.

For this experiment, we measure how quickly IDOLpro can generate a ligand with both better binding affinity (Vina score) and synthetic accessibility (SA score) than the ligand with the best binding affinity found when screening ZINC250K. For screening ZINC250K, we use QuickVina2~\cite{alhossary2015fast} to quickly dock molecules to each of the 10 target protein pockets in the RGA test set. We run docking on an AWS compute-optimized instance with 8 CPU cores. We then use IDOLpro to generate optimized ligands, making note of the number of ligands, time, and cost required for IDOLpro to generate an improved ligand relative to virtually screening ZINC250K. Cost is based on the AWS pricing for the requested instances. The results are shown in~\cref{tab:vs_compare}. 

Screening ZINC250K using QuickVina2 takes an average of $\approx 161$ hours per protein pocket. On the other hand, IDOLpro is able to find a ligand with a better Vina and SA score than the virtual screen in under an hour ($\approx 56$ minutes) on average. This translates to $\approx 173\times$ speedup in terms of time, and $\approx 103\times$ reduction in cost. For 4/10 cases, IDOLpro is able to find a ligand with better Vina and SA score within the first 10 ligands generated, and for 9/10 cases within the first 100 ligands generated. For a single case (PDB ID 3eml) IDOLpro needs to produce 153 ligands to find a ligand with better binding affinity and synthetic accessibility than the one found in the virtual screen. Even in this case, running IDOLpro takes $\approx 2.5$ hours to run on the requested GPU instance, translating to a $\approx 66\times$ speedup in terms of time, and $\approx 39\times$ reduction in cost compared to virtually screening ZINC250K.

In general, for a given protein pocket, IDOLpro produces ligands with significantly better binding affinities than those found when screening a standard drug database such as ZINC250K. In \cref{fig:vs_vs_idolpro}, we track the average running top-1 and top-10 Vina scores for IDOLpro across the 10 protein targets as a function of the number of ligands generated. For both the top-1 and top-10 Vina score, IDOLpro quickly surpasses the virtual screen of ZINC250K -- needing to generate on average only 1 ligand to surpass the top-1 Vina score of the virtual screen, and needing to generate on average only 15 ligands to surpass the top-10 Vina score of the virtual screen. We also plot the distribution of Vina scores for both the ligands produced by IDOLpro, and those found when screening ZINC250K for 2 of the targets from the RGA test set -- 2rgp, a protein whose over-expression has been associated with human tumor growth~\cite{xu20084}, and 7l11 -- the SARS-CoV-2 main protease~\cite{zhang2021potent}. These plots are shown in \cref{fig:vs_vs_idolpro}. These plots further validate IDOLpro's ability to generate ligands with high binding affinity relative to virtual screening.

\begin{table*}[ht!]
\centering
\begin{tabular}{c|c|c|c|}
    Method & $N_{\text{ligands}}$  & Time [h] & Cost [\$] \\ \hline 
    IDOLpro & $44.90 \pm 46.06$ & $0.93 \pm 0.75$ & $1.06 \pm 0.86$ \\
    Virtual Screen & $250,000$ & $160.90 \pm 24.12$ & $109.41 \pm 16.40$ 
\end{tabular}
\caption{Comparison of using IDOLpro to find improved ligands relative to a virtual screen of ZINC250K. We note the average number of ligands, time, and cost it takes for IDOLpro to find a ligand with both better Vina and SA than the best ligand from ZINC250K. Virtual screening was run on an AWS compute-optimized instance with 8 CPU cores, while IDOLpro was run on an NVIDIA A10G GPU with 24 GB of VRAM. Costs are computed based on AWS instance pricing.}
\label{tab:vs_compare}
\end{table*}

\subsection{Comparison to Other Methods}\label{subsec:compare-ga}

Lastly, we compare IDOLpro to various non-deep learning-based methods in the literature. These methods include genetic algorithms~\cite{nigam2019augmenting,jensen2019graph,spiegel2020autogrow4,fu2022reinforced}, reinforcement learning~\cite{zhou2019optimization,ahn2020guiding,olivecrona2017molecular,jin2020multi}, and and an MCMC method~\cite{xie2021mars}. We evaluate these methods across the 10 protein pockets in the RGA test set. For each target, as was done in Ref.~\cite{fu2022reinforced}, we generate 1000 ligands with IDOLpro and calculate the average top-100, top-10, and top-1 Vina score. We also record the average SA and QED of molecules, along with the average diversity per protein pocket. Numbers for other methods are taken from Ref~\cite{fu2022reinforced}. Results are shown in~\cref{tab:ga_results}.

IDOLpro greatly outperforms non-DL techniques in terms of Vina score, improving on the next best method by $\approx23\%$, $\approx 29\%$, and $\approx35\%$ in terms of average top-100, top-10, and top-1 Vina score respectively. Unlike when compared to other DL methods, IDOLpro ranks behind most of these methods in terms of average SA, ranking 9th out of the 10 methods compared. IDOLpro is middle-of-the-pack in terms of QED, ranking 5th out of the 10 methods compared. This shows that IDOLpro, and deep learning methods in general, have a ways to go before they can produce molecules with the same synthesizability and drug-likeness as other advanced non-DL methods in the literature. This is an ongoing area of research~\cite{gao2020synthesizability,buttenschoen2024posebusters}, and is an aspect that we would like to improve in IDOLpro.

\begin{table*}[ht!]
\centering
\begin{tabular}{c|c|c|c|c|c|c}
    Method & Vina$_{\text{top-100}}$ [kcal/mol] & Vina$_{\text{top-10}}$ [kcal/mol] & Vina$_{\text{top-1}}$ [kcal/mol] & SA & QED & Diversity \\ \hline
    MARS~\cite{xie2021mars} & $-7.76 \pm 0.61$ & $-8.8 \pm 0.71$ & $-9.26 \pm 0.79$ & $2.69 \pm 0.08$ & $0.71 \pm 0.01$ & $\mathbf{0.88 \pm 0.00}$ \\
    MolDQN~\cite{zhou2019optimization} & $-6.29 \pm 0.40$ & $-7.04 \pm 0.49$ & -$7.50 \pm 0.40$ & $5.83 \pm 0.18$ &  $0.17 \pm 0.02$ & $\mathbf{0.88 \pm 0.01}$ \\
    GEGL~\cite{ahn2020guiding} & $-9.06 \pm 0.92$ & $-9.91 \pm 0.99$ & $-10.45 \pm 1.04$ & $2.99 \pm 0.05$ & $0.64 \pm 0.01$ & $ 0.85 \pm 0.00$ \\
    REINVENT~\cite{olivecrona2017molecular} & $-10.81 \pm 0.44$ & $-11.23 \pm 0.63$ & $-12.01 \pm 0.83$ & $2.60 \pm 0.12$ & $0.45 \pm 0.06$ & $0.86 \pm 0.01$ \\
    RationaleRL~\cite{jin2020multi} & $-9.23 \pm 0.92$ & $-10.83 \pm 0.86$ & $-11.64 \pm 1.10$ & $2.92 \pm 0.13$ & $0.32 \pm 0.02$ & $0.72 \pm 0.03$  \\
    GA+D~\cite{nigam2019augmenting} & $-8.69 \pm 0.45$ & $-9.29 \pm 0.58$ & $-9.83 \pm 0.32$ & $3.45 \pm .12$ &  $0.70 \pm 0.02$ & $0.87 \pm 0.01$ \\
    Graph-GA~\cite{jensen2019graph} & $-10.48 \pm 0.86$ & $-11.70 \pm 0.93$ & $-12.30 \pm 1.91$ & $3.50 \pm 0.37$ & $0.46 \pm 0.07 $ & $0.81 \pm 0.04$ \\
    Autogrow 4.0~\cite{spiegel2020autogrow4} & $-11.37 \pm 0.40$ & $-12.21 \pm 0.62$ & $-12.47 \pm 0.84$ & $2.50 \pm 0.05$ & $\mathbf{0.75 \pm 0.02}$ & $0.85 \pm 0.01 $ \\
    RGA~\cite{fu2022reinforced} & $-11.87 \pm 0.17$ & $-12.56 \pm 0.29$ & $-12.89 \pm 0.47$ & $\mathbf{2.47 \pm 0.05}$ & $0.74 \pm 0.04 $ & $0.86 \pm 0.02$ \\
    IDOLpro & $\mathbf{-14.59 \pm 1.51}$ & $\mathbf{-16.26 \pm 1.66}$ & $\mathbf{-17.35 \pm 2.10}$ & $3.77 \pm 0.33$ & $0.64 \pm 0.06 $ & $0.72 \pm 0.04$ \\
    
\end{tabular}
\caption{Comparison of IDOLpro to various score and sample-based methods on 10 disease-related protein targets. The average top-100, top-10, and top-1 Vina scores across the targets are reported, along with the average SA, average QED, and diversity. The top-performing model on each metric is bolded in the corresponding column. Numbers for other methods are taken from Ref.~\cite{fu2022reinforced}.
}
\label{tab:ga_results}
\end{table*}

\subsection{Lead Optimization}\label{subsec:lead-opt}

In addition to \textit{de novo} generation, IDOLpro can be used to optimize a known ligand in the protein pocket. This functionality is useful for a common task in drug discovery pipelines known as lead optimization, where a molecule is progressed from an initial promising candidate towards having optimal properties~\cite{hughes2011principles}. Generally, this is accomplished by fixing a large part of the molecule, i.e., the scaffold, while optimizing the rest of the molecule~\cite{bohm2004scaffold}. This is incorporated into generation with DiffSBDD~\cite{schneuing2022structure} using an inpainting-inspired approach. At each denoising time step, fixed atoms are replaced with their noised counterparts, while the rest of the molecule is generated \textit{de-novo}.

We test the capability of our framework for performing lead optimization with examples from the CrossDocked test set. We specify atoms to fix by using RDKit to identify the the Bemis-Murcko scaffold~\cite{bemis1996properties} of each reference ligand. If no scaffold is found (13 cases), if it is identified to be greater than $90\%$ of the full reference ligand (13 cases), or if it contains atoms not supported by the ANI2x model (3 cases), then these targets are removed from the test set. The lead optimization results with IDOLpro for the remaining 71 protein and scaffold pairs are shown in~\cref{tab:ref_opt_improve} where they are compared to the original reference ligands. We report the Vina and SA scores of the un-docked reference ligands since their scaffolds' coordinates act as the seed for IDOLpro optimization.  

We find that the average Vina scores of the optimized ligands exceed the average values of the reference ligands in the test set (1.59 kcal/mol). Although the average SA is higher, both the top-10\% SA score and the top-10\% Vina are significantly better than the test set, and are more realistic to consider when assessing lead optimization capability. A molecule with both better SA and Vina scores compared to the reference ligands was found for all but one target, where we were able to improve the SA score but not the Vina score. As discussed previously, a possible remedy would be re-weighting the optimization objective to more heavily favor Vina score.  

Two examples of generated molecules that retain the original seed scaffold and successfully improve both the Vina and SA scores compared to the reference ligand are shown in~\cref{fig:ref_opt_examples}. The first is a case where the identified scaffold is a small portion of the reference (ligand with PDB residue ID ily docked into protein 1a2g), and we see a large diversity in the generated molecules. The second is a case where the scaffold is a large component of reference (ligand with PDB residue ID plp docked into protein 2jjg) where we are effectively optimizing functional groups on a fixed scaffold. We believe our results demonstrate the flexibility of atom fixing in IDOLpro, and its utility in performing lead optimization.

\section{Discussion}
\label{conclusion}

\begin{figure*}[ht!]
    \centering
        \includegraphics[width=0.75\linewidth]{figs/5.pdf}
    \caption{Molecules produced by IDOLpro during lead optimization. Examples shown are on the ligand 1ly docked into protein 1a2g, and ligand plp docked into protein 2jjg, both examples from the CrossDocked test set. IDOLpro is used to append molecules to the scaffold while optimizing torchvina and torchSA. IDOLpro yields multiple ligands with improved SA and Vina relative to the reference molecule.} 
    \label{fig:ref_opt_examples}
\end{figure*}

\begin{table*}[ht!]
\centering
\begin{tabular}{c|c|c|c|c|}
    Method & Vina [kcal/mol] & Vina$_{10\%}$ [kcal/mol] &  SA & SA$_{10\%}$   \\ \hline 
    Test Set &  -5.58 $\pm$ 2.32 & - & 3.67 $\pm$ 1.23 & -   \\
    IDOLpro &  -7.17 $\pm$ 2.36 & -8.96 $\pm$ 2.57 & 4.12 $\pm$ 1.10 & 2.90 $\pm$ 1.12  \\
\end{tabular}
\caption{Results for IDOLpro scaffold fixing on the 71 crossdocked data points with identifiable scaffolds using RDKit's Bemis-Murcko scaffold. The average and top-10 Vina and SA scores are reported for the reference ligands and the optimized ligands.}
\label{tab:ref_opt_improve}
\end{table*}

We have presented a framework that is designed to produce optimal ligands with desired properties for a given protein pocket. We accomplish this by constructing a computational graph that begins with the latent variables of a diffusion model and ends with the evaluation of metrics important in drug discovery. The latent variables can then be modified via standard gradient-based optimization routines to optimize the metrics of interest. More specifically, we perform multivariate optimization by optimizing both Vina and SA scores simultaneously. The molecules generated by our platform achieve the best Vina scores when compared to previous, state-of-the-art machine learning methods. When considering the CrossDocked and Binding MOAD test sets, we see a 10\% (0.71 kcal/mol) and 20\% (1.43 kcal/mol) improvement to the next best tool. Our tool ranks second among all tools compared for producing molecules with good SA scores on CrossDocked, and is state-of-the-art for the Binding MOAD, improving upon the SA of the next best tool by 35\%. Furthermore, our tool produces molecules with the highest QED on both datasets. For the Binding MOAD, our framework is the first to produce molecules with a better average Vina score than reference molecules in the test set, which were derived through experiment. Our tool shows great promise in the hit-finding stage of a typical drug discovery pipeline, finding molecules with better Vina and SA scores at a fraction of the time and cost when compared to a virtual screen of a library of commercially available compounds. In addition, our tool can perform lead optimization by beginning the optimization with a scaffold taken from a reference molecule. When applied to a relevant subset of the CrossDock dataset, we identified ligands with both better SA and Vina scores than the reference ligand in all but one case. Our framework proposes optimal drugs given particular properties, unlocking the ability to virtually screen optimized molecules from a vast chemical space without the need of searching through a large database, or generating molecules randomly until one with desired properties is found, therefore accelerating the drug discovery process. Our future work will include other important metrics such as toxicity and solubility within the objective to ensure the generation of feasible ligands, along with the consideration of other binding affinity metrics such as free energy perturbation (FEP) based affinity~\cite{crivelli2023machine}.

\section{Methods}
\label{methods}

\subsection{Generator Module}

When optimizing latent vectors, we utilize a state-of-the-art denoising diffusion probabilistic model (DDPM)~\cite{ho2020denoising}, DiffSBDD \cite{schneuing2022structure}, for generating novel ligands with high binding affinity. DDPMs generate samples from a target distribution by learning the a denoising process. For some $\Tilde{\mathbf{z}}_{data}$ sampled from the target distribution, a diffusion process adds noise to $\Tilde{\mathbf{z}}_{data}$ to elicit the latent vector at time-step $t$ for $t = 0,...,T$ (where $T$ is the length of the noising process) according to the transition distributions defined by:
\begin{align}\label{eq:noising}
    p(\mathbf{z}_t | \Tilde{\mathbf{z}}_{data}) = \mathcal{N}(\mathbf{z}_t | \alpha_t \Tilde{\mathbf{z}}_{data}, \sigma_t^2 \mathbf{I}),
\end{align}
$\alpha_t$ controls the level of noise in $\mathbf{z}_t$ and follows a pre-defined schedule from $\alpha_0 \approx 1$ (no noise) to $\alpha_T \approx 0$ (pure noise). In DiffSBDD, the variance-preserving noising process is used, i.e., $\alpha_t = \sqrt{1 - \sigma_t^2}$. The posterior of the transitions conditioned on $\Tilde{\mathbf{z}}_{data}$ define the inverse of the noising process, i.e., the denoising process, and can be written in closed form for any $s < t$:
\begin{align}\label{eq:denoising}
    q(\mathbf{z}_s | \Tilde{\mathbf{z}}_{data}, \mathbf{z}_t) = \mathcal{N}\left(\mathbf{z}_s | \frac{\alpha_{t|s}\sigma_s^2}{\sigma_t^2}\mathbf{z}_t + \frac{\alpha_s \sigma^2_{t|s}}{\sigma_t^2}\Tilde{\mathbf{z}}_{data}, \frac{\sigma_{t|s}^2\sigma_s^2}{\sigma_t^2}\mathbf{I} \right),
\end{align}
where $\alpha_{t|s} =\frac{\alpha_t}{\alpha_s}$, and $\sigma_{t|s} = \sigma_t^2 - \alpha^2_{t|s}\sigma_s^2$ following the notation of Ref.~\cite{hoogeboom2022equivariant}. This denoising process relies on $\Tilde{\mathbf{z}}_{data}$, which is not available during inference. Instead, a neural network, $\psi_{\theta}$, is used to make an inference of $\mathbf{\hat{z}}_{data}$. Ref.~\cite{ho2020denoising} found that optimization is easier when predicting the noise in the signal. One can reparameterize Eq.~\eqref{eq:noising} such that $\Tilde{\mathbf{z}}_{data} = \frac{1}{\alpha_t}\mathbf{z}_t - \frac{\sigma_t}{\alpha_t} \epsilon$ where $\epsilon \sim \mathcal{N}(\mathbf{0},\mathbf{I})$, and get the neural network to predict $\hat{\epsilon}_{\theta} (\mathbf{z}_t, t)$, yielding a prediction $\hat{\mathbf{z}}_{data}$. Thus, given $\mathbf{z}_t$ and using Eq.~\eqref{eq:denoising}, one can sample $\mathbf{z}_s$ for any $s < t$. In practice, denoising is done in successive time-steps, i.e., $s = t-1$ in Eq.~\eqref{eq:denoising}.

DiffSBDD is an \textit{SE(3)}-equivariant~\cite{fuchs2020se} 3D-conditional DDPM which respects translation, rotation, and permutation symmetries. DiffSBDD was trained to create ligands with high binding affinity given a target protein pocket. In DiffSBDD, data samples consist of protein pocket and ligand point clouds, i.e., $\mathbf{z} = [\mathbf{z}^{\ell}, \mathbf{z}^p]$. Each point cloud consists of atom types and coordinates, $\mathbf{z} = [\mathbf{r}, \mathbf{h}]$ where $\mathbf{r} \in \mathbb{R}^{N \times 3}$ is a tensor of atomic coordinates, and $\mathbf{h} \in \mathbb{R}^{N \times 10}$ is a tensor of atomic probabilities over the atom types which the model can generate. Within the model, each $\mathbf{z}_t$ is converted to a graph, and processed by an EGNN~\cite{satorras2021n} to produce a prediction $\hat{\epsilon}_{\theta} (\mathbf{z}_t, t)$. DiffSBDD contains two different models for 3D pocket conditioning -- a conditional DDPM that receives a fixed pocket representation as the context in each denoising step, and a model that is trained to approximate the joint distribution of ligand-protein pocket pairs and is combined with an inpainting-inspired~\cite{song2020score,lugmayr2022repaint} sampling procedure at inference time to yield conditional samples of ligands given a fixed protein pocket. In this work, we focus only on the conditional DDPM for ligand generation. Our framework requires that generated ligands do not overlap with the target protein pocket, as the ligands' poses later undergo local structural refinement. We found that the conditional DDPM model can consistently generate ligands satisfying this constraint.

For each pocket-conditioning scheme, DiffSBDD contains two versions of the model - one trained on a subset of CrossDocked~\cite{francoeur2020three}, and another trained on a subset of the Binding MOAD ~\cite{hu2005binding}. For CrossDocked, Ref.~\cite{schneuing2022structure} used the same train/test splits as in Ref.~\cite{luo20213d} and Ref.~\cite{peng2022pocket2mol}, resulting in 100,000 complexes for training, and 100 protein pockets for testing. For the Binding MOAD, Ref.~\cite{schneuing2022structure} filtered the database to contain only molecules with atom types compatible with their model, and removed corrupted entries, resulting in 40,344 complexes for training, and 130 protein pockets for testing. In this work we use only version of DiffSBDD trained on CrossDocked as we found we were able to achieve good results on both test sets (CrossDocked and the Binding MOAD) with just this model.

DiffSBDD was shown to achieve state-of-the-art performance on both test sets. In particular, DiffSBDD achieved the best average, and best top-10\% Vina score when compared with other state-of-the-art models in the literature -- 3D-SBDD~\cite{luo20213d}, Pocket2Mol\cite{peng2022pocket2mol}, GraphBP~\cite{graphbp}, and TargetDiff~\cite{targetdiff}. We note, that although DiffSBDD is used as our baseline model in this report, our framework is not limited to the use of this specific model -- any other diffusion model can take its place.

\begin{figure*}[ht]
    \centering
    \begin{subfigure}{0.45\textwidth}
        \centering
        \includegraphics[width=\linewidth]{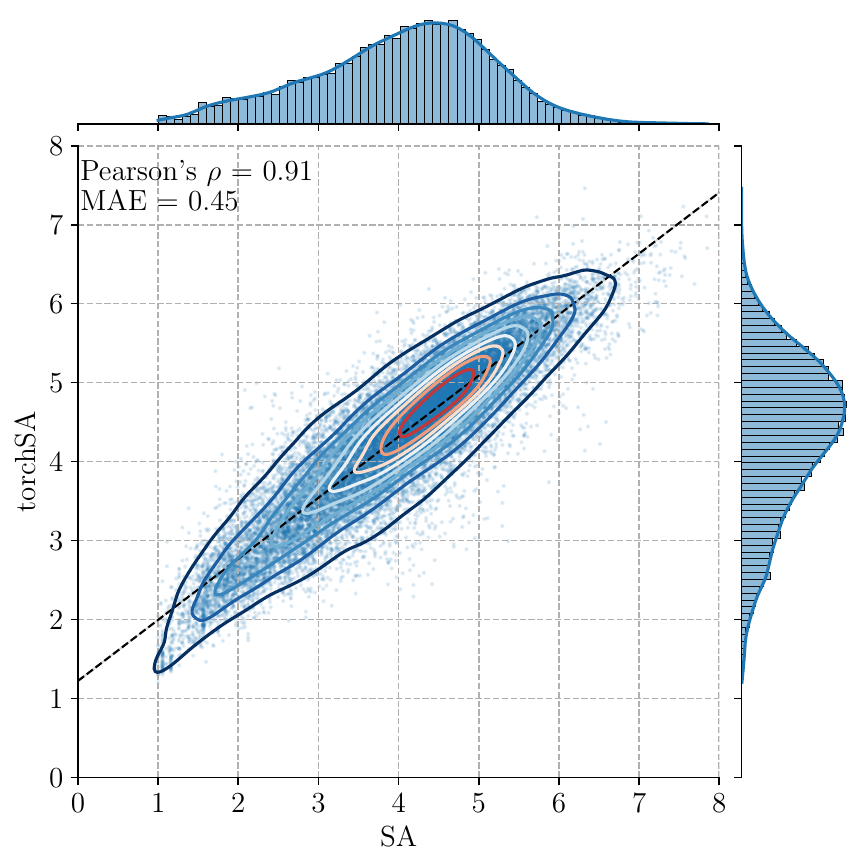}
    \end{subfigure}
    \caption{Correlation plot between the predictions of the torchSA model and the SA score as computed in RDKit~\cite{rdkit}. The torchSA model is validated on $\approx 25,000$ atomic point clouds (coordinates and atomic probability distributions) generated by DiffSBDD. For each atomic point cloud, we construct a corresponding RDKit molecule to assess the SA score. TorchSA displays a high-correlation ($\rho = 0.91$) and low mean absolute error ($MAE=0.45$) relative to the SA score on these data points. The histogram on each axis represents the distribution of molecules that achieved corresponding scores as predicted by torchSA (y-axis) and SA score (x-axis).}
    \label{fig:torchsa_validation}
\end{figure*}

\subsection{Molecule size}
\label{subsec:size}
For determining the number of atoms in a generated ligand $n_{\ell}$, we employ the same sampling procedure as is used when sampling structures from DiffSBDD~\cite{schneuing2022structure}. We sample from an empirical distribution of the number of heavy atoms in the ligand given the number of heavy atoms in the pocket -- $p(n_{\ell}|n_p)$. This empirical distribution is based on the data used to train the CrossDocked model in Ref.~\cite{schneuing2022structure}. We then bias the model in the same way that is done in Ref.~\cite{schneuing2022structure}, adding 5 heavy atoms to each sample of $n_{\ell}$ -- this allows us to have an apples-to-apples comparison of our tool versus tools compared in that work. In lead optimization, the number of heavy atoms in the ligand is sampled from a normal distribution centered at the size of the reference ligand, and whose lower limit is clamped to be no lower than the number of heavy atoms in the scaffold. The number of atoms in the scaffolds relative to their reference values varied in the test set. To ensure a balanced sampling of scaffolds with more or fewer atoms than the reference, in this study we set the standard deviation to the greater of 5 (matching the heavy atom bias) or half the difference between the number of heavy atoms in the scaffold and the reference ligand.

\subsection{Ligand Validity Checks}
\label{subsec:validity}
When generating ligands using DiffSBDD, we perform several chemical and structural checks to ensure that the generated ligand is valid. A number of these checks use RDKit~\cite{rdkit}. These include a valency check, verifying that hydrogens can be added to the ligand and assigned a Cartesian coordinate (using the \textit{addCoords} option in \textit{Chem.AddHs}), that the ligand is not fragmented, and that the ligand can be sanitized. All of these except for the valency check are also done within DiffSBDD~\cite{Arne2013}. 

In addition, we have four more checks to ensure the structural validity of the ligand. These four checks are necessary to be able to run structural refinement with IDOLpro, which makes use of the ANI2x model~\cite{devereux2020extending}. Structural refinement is described in \cref{subsec:docking}. We first make sure that the ligand contains only atoms compatible with ANI2x. DiffSBDD can generate ligands with four atom types that are incompatible with ANI2x -- B, P, Br, and I. We also make sure that the bond lengths in the ligand are reasonable by referring to covalent radii, and that the ligand does not overlap with the protein pocket. This is done via ASE's~\cite{larsen2017atomic} (Atomic Simulation Environment) \textit{NeighborList} class. Lastly, We make sure that the atoms do not have significant overlap within the ligand itself. This is done via pymatgen's~\cite{ong2013pymatgen} \textit{Molecule} class.

\subsection{Scoring Module}
After generating a set of ligands, we pass them to a scoring module. In this work, we include a custom torch-based Vina score~\cite{trott2010autodock} which we refer to as torchvina, an equivariant neural network trained to predict the synthetic accessibility of molecules with 3D information~\cite{ertl2009estimation} which we refer to as torchSA, and the ANI2x model~\cite{devereux2020extending}. These objectives are all written using Pytorch~\cite{paszke2019pytorch} with differentiable operations and hence can be differentiated automatically using autograd. \\

\noindent\textbf{torchvina}\; We re-implement the Vina force field~\cite{trott2010autodock} using Pytorch to allow for automatic differentiation with respect to the latent parameters of the generator. Our work is not the first to produce a Pytorch-based version of Vina to facilitate automatic differentiation, a similar implementation was presented by Ref.~\cite{wang2023fully}. Our motivation for implementing a differentiable Vina score is that docking with Vina was shown to outperform state-of-the-art ML models such as DiffDock~\cite{corso2022diffdock} when stricter chemical and physical validity checks were enforced on docked molecules, or when these procedures were evaluated on a dataset composed of examples distinct from the ML models' training data~\cite{buttenschoen2024posebusters}.

The Vina force field is composed of a weighted sum of atomic interactions. Steric, hydrophobic, and hydrogen bonding interactions are calculated and weighted according to a nonlinear fit to structural data~\cite{trott2010autodock}. The final score is re-weighted by the number of rotatable bonds to account for entropic penalties~\cite{mcnutt2021gnina}. The Vina score is composed of a sum of intramolecular and intermolecular terms, both of which are integrated into our implementation.

To validate our implementation of torchvina, we took 4 systems from our validation set (details in the SI). The Vina score consists of a weighted sum of five individual terms, three steric terms -- two attractive Gaussian terms (gauss1 and gauss2), and a repulsion term, as well as hydrophobic interactions, and where applicable, hydrogen bonding. For these four systems, we made sure that torchvina produced the same value on each of these sub-terms, as well as the total intermolecular energy and the total intramolecular energy as the smina~\cite{koes2013lessons} tool. These checks are included in our github repository in the tests directory. \\

\noindent\textbf{torchSA}\; To have an evaluator model capable of estimating synthesizability, we train an equivariant neural network to predict the synthetic accessibility (SA) score. SA score was first proposed by Ref.~\cite{ertl2009estimation}, ranges from 1 (easy to make) and 10 (very difficult to make), and shown to be effective for biasing generative pipelines towards synthesizable molecules~\cite{gao2020synthesizability,skoraczynski2023critical}. Moreover, it was used directly in DiffSBDD to measure the performance of the pipeline~\cite{schneuing2022structure}. To be able to guide latent parameters in DiffSBDD towards generating ligands with high synthesizability requires designing a model that can handle the outputs of DiffSBDD in a differentiable manner. In particular, we construct a machine learning model that can take in atomic point clouds, $\mathbf{z} = [\mathbf{r}, \mathbf{h}]$, where $\mathbf{r}$ is a set of coordinates, and $\mathbf{h}$ is corresponding probability distributions over atom types.
We train this model by creating a dataset of atomic point clouds of ligands labeled with SA score. To allow for predictions on probability distributions of atom types, we encode atom types as one-hot vectors. For more details on the training of this model, we refer the reader to the SI.

\begin{algorithm}[ht!]
\caption{Optimization with IDOLpro}\label{alg:cap}
\begin{algorithmic}
\Variables
\State $N_{max}$, maximum number of optimization steps.
\State $\Tilde{\mathbf{z}}^p$, protein atom embeddings (types and coordinates).
\State $n_{\ell}$, number of atoms in generated ligand.
\State $\Tilde{\mathbf{z}}_0^{\ell}$, fixed ligand atom embeddings (for lead optimization).
\State $\mathbf{m}_{\ell}$, mask for fixed ligand atoms (for lead optimization).
\EndVariables
\State
\Function{IDOLpro}{$\Tilde{\mathbf{z}}^p$, $n_{\ell}$, $\Tilde{\mathbf{z}}_0^{\ell}$ = None, $\mathbf{m}_{\ell} = \mathbf{1}$}
\State Sample $\mathbf{z}^{\ell}_T  \sim \mathcal{N}(\mathbf{0}_{n_{\ell}}, \mathbf{I}_{n_{\ell}})$
\For{$t=T,...,t_{hz}+1$}
    \State $\mathbf{z}^{\ell}_{t-1} \gets$ \Call{Denoise}{$\mathbf{z}^{\ell}_{t}, \Tilde{\mathbf{z}}^{p}, \mathbf{m}_{\ell}, \Tilde{\mathbf{z}}_0^{\ell}$}
\EndFor
\For{$i=0,...,N_{max}$}
    \For{$t=t_{hz},...,1$}
        \State $\mathbf{z}^{\ell}_{t-1} \gets$ \Call{Denoise}{$\mathbf{z}^{\ell}_{t}, \Tilde{\mathbf{z}}^p, \Tilde{\mathbf{z}}_0^{\ell}, \mathbf{m}_{\ell}$}
    \EndFor
    \If{\Call{Validity}{$\mathbf{z}^{\ell}_0, \Tilde{\mathbf{z}}^p$}}
        \State $\mathbf{z}^{\ell}_{t_{hz}} \gets$ \Call{Adam}{$\mathbf{z}^{\ell}_{t_{hz}}, \frac{\partial (\text{\texttt{Vina}}(\mathbf{z}^{\ell}_0, \Tilde{\mathbf{z}}^p) + \text{\texttt{SA}}(\mathbf{z}^{\ell}_0))}{\partial \mathbf{z}^{\ell}_{t_{hz}}}$}
    \EndIf
\EndFor
\For{$t=t_{hz},...,1$}
    \State $\mathbf{z}^{\ell}_{t-1} \gets$ \Call{Denoise}{$\mathbf{z}^{\ell}_{t}, \Tilde{\mathbf{z}}^p,  \Tilde{\mathbf{z}}_0^{\ell}, \mathbf{m}^{\ell}$}
\EndFor
\State $\mathbf{z}^{\ell}_0 \gets$ \Call{StructuralRefinement}{$\mathbf{z}_0^{\ell}$}
\State \Return $\mathbf{z}^{\ell}_0$
\EndFunction
\State
\Function{Validity}{$\mathbf{z}^{\ell}, \mathbf{z}^{p}$}
    \State $v \gets$ \Call{ValenceCheck}{$\mathbf{z}^{\ell}$}
    \State $v \gets v$ and \Call{CanSanitize}{$\mathbf{z}^{\ell}$}
    \State $v \gets v$ and \Call{CanAddHs}{$\mathbf{z}^{\ell}$}
    \State $v \gets v$ and \Call{Fragments}{$\mathbf{z}^{\ell}$} = 1
    \State $v \gets v$ and $h^{\ell}_i \in \{H, C, N, O, F, Cl, S\}$
    \State $v \gets v$ and \Call{Connected}{$\mathbf{z}^{\ell}$}
    \State $v \gets v$ and \Call{NoOverlapIntra}{$\mathbf{z}^{\ell}$}
    \State $v \gets v$ and \Call{NoOverlapInter}{$\mathbf{z}^{\ell}, \mathbf{z}^{p}$}
    \State \Return $v$
\EndFunction
\State
\Function{Denoise}{$\mathbf{z}^{\ell}_t, \Tilde{\mathbf{z}}^{p}, \mathbf{z}_0^{\ell}, \mathbf{m}_{\ell}$}
    \If{$\mathbf{z}_0^{\ell}$ is None} \Comment{de-novo}
        \State $\mathbf{z}^{\ell}_{t-1} \gets q(\mathbf{z}_{t-1}^{\ell} | \mathbf{z}_{t}^{\ell}, \mathbf{z}^p)$
    \Else \Comment{lead-opt}
        \State $\mathbf{z}^{\ell}_{t-1} \gets q(\mathbf{z}_{t-1}^{\ell} | \mathbf{z}_{t}^{\ell}, \mathbf{z}^{p}) \cdot \mathbf{m}_{\ell} + p(\mathbf{z}^{\ell}_{t-1} | \mathbf{z}^{\ell}_0, \mathbf{z}^{p}) \cdot (1 - \mathbf{m}_{\ell})$
    \EndIf
\EndFunction
\end{algorithmic}
\end{algorithm}

To validate the torchSA model for guiding DiffSBDD towards generating molecules with high synthetic accessibility, we generated 100 molecules using IDOLpro with only torchvina guidance for each target in our validation set (see SI for details on structures in this set). For each generated molecule, we save the entire trajectory of point clouds required to produce the final IDOLpro molecule, i.e., the point cloud produced by DiffSBDD during each gradient update of the latent vectors. For each of these intermediate point clouds, we store the coordinates, and raw atomic probabilities. For each  point cloud, we also store their corresponding RDKit molecule. This is done by taking the most likely atom type for each probabilty distribution, and adding bonds using OpenBabel~\cite{o2011open}. We generate 25,926 individual point cloud-molecule pairs for evaluating the torchSA model. For each pair, we compare the output of torchSA on the point cloud to the SA score on the corresponding molecule. We plot the correlation between the two in \cref{fig:torchsa_validation}, and include both the Pearson's $\rho$, and mean absolute error in the figure. \\

\begin{table*}[ht!]
\centering
    \begin{tabular}{c|c|c|c|c|c|c|c}
         Method & Torchvina & ANI2x & $L1$ bond penalty &  Vina [kcal/mol] &  Vina$_{10\%}$ [kcal/mol] & Validity [\%] & Time [s] \\\hline
         QuickVina2 & - & - & - & -8.51 & -9.51 & 95.0 & 75.0 \\
         IDOLpro & Inter+Intra & Inter+Intra & \xmark & -9.26  & -10.35  & 80.1  & 19.34\\
         IDOLpro & Inter+Intra & Intra & \xmark & -9.33 & -10.41 & 77.2 & 22.49 \\
         IDOLpro& Inter & Inter+Intra & \xmark & -9.38 & -10.53 & 82.1 & 18.53 \\
         IDOLpro &Inter & Intra & \xmark &  -9.53 & -10.70 & 81.6 & 23.90 \\
         IDOLpro& Inter & Intra & \cmark &  -9.39 & -10.68 & 86.6 & 24.45 \\
    \end{tabular}
    \caption{Results when running structural refinement with various combinations of inter- and intra-molecular forces derived from the Vina and ANI2x potentials. Additionally, we note whether an $L1$ penalty for violating bonds was used. For each experiment, we report the Vina score, top-10 Vina score, and the percent of valid structures produced. Validity checks are performed according to the checks described in \cref{subsec:validity}. Lastly, we report the average time to run structural refinement on the 20 ligands generated during each run. Experiments were run on an NVIDIA A10G GPU with 24 GB of memory. }
\label{tab:structural-refine}
\end{table*}

\noindent\textbf{ANI2x}\; ANI2x is a neural network ensemble model that is part of the ANI suite of models~\cite{torchani}. The ANI models are trained on quantum chemistry calculations (at the density functional theory level) and they predict the total energy of a target system.
The ANI models are trained on millions of organic molecules and are accurate across different domains~\cite{smith2017ani,smith2018less,devereux2020extending,smith2019approaching}. In addition, they have been shown to outperform many common force fields in terms of accuracy \cite{folmsbee2021assessing}. The ANI models make use of atomic environment descriptors, which probe their local environment, as input vectors. An individual ANI model contains multiple neural networks, each specialized for a specific atom type, predicting the energy contributed by atoms of that type in the molecular system. The total energy of the system is obtained by performing a summation over the atomic contributions~\cite{smith2017ani}. The ANI2x model is an ensemble model consisting of 8 individual ANI models. Each sub-model is trained on a different fold of the ANI2x dataset, composed of gas-phase molecules containing seven different atom types -- H, C, N, O, F, Cl, and S~\cite{devereux2020extending}. These seven atom types cover $\approx 90\%$ of drug-like molecules, making ANI2x a suitable ML model for usage in our framework.

\subsection{Latent Vector Optimization}

The main optimization in IDOLpro occurs via the modification of latent vectors used by the generator to generate novel ligands. We do this by repeatedly evaluating generated ligands with an objective composed of a sum of differentiable scores, calculating the gradient of the objective with respect to the latent vectors (facilitated by automatic differentiation with Pytorch~\cite{paszke2019pytorch}), and modifying the latent vectors via a gradient-based optimizer. 

When optimizing latent vectors in DiffSBDD, we do not modify the initial latent vectors used by the model. Instead, we define an optimization horizon, $t_{\text{hz}}$. First latent vectors are generated up to the optimization horizon $\mathbf{z}^{\ell}_{T}, \hdots, \mathbf{z}^{\ell}_{t_{\text{hz}}}$. This latent vector is saved, and the remaining latent vectors, $\mathbf{z}^{\ell}_{t_{\text{hz}}-1}, \hdots, \mathbf{z}^{\ell}_{0}$, are generated. Upon passing the ligand validity checks (Section~\ref{subsec:validity}), the gradient of the objective with respect to $\mathbf{z}_{t_{\text{hz}}}$ is evaluated, and $\mathbf{z}_{t_{\text{hz}}}$ is modified using a gradient-based optimizer. When re-generating ligands, rather than starting from $\mathbf{z}^{\ell}_T$, only latent vectors proceeding the optimization horizon are re-generated, i.e., $\mathbf{z}^{\ell}_{t_{\text{hz}}-1}, \hdots, \mathbf{z}^{\ell}_{0}$. Optimization continues until a maximum number of steps, $N_{max}$, have been taken in the latent space. When de-noising $\mathbf{z}^{\ell}_t$ during lead optimization, a mask $\mathbf{m}_{\ell} \in \{0, 1\}^{n_{\ell}}$ is used to keep specific atoms fixed during the inpainting procedure. We employ backtracking, early-stopping, and learning-rate decay, all described in the SI. The optimization of a single ligand is provided in Algorithm 1.

In this work, we focus on using two combinations of evaluators: torchvina on its own, and torchvina in combination with torchSA. We use the Adam~\cite{kingma2014adam} optimizer with a learning rate of $0.1$, $\beta_1=0.5$ and $\beta_2=0.999$ to modify latent vectors. We perform hyperparameter optimization to choose the optimization horizon, described in the SI.

\subsection{Structural Refinement}\label{subsec:docking}

After an optimized ligand has been generated into a protein pocket with latent vector optimization, its bound pose is further refined via structural refinement. Structural refinement in IDOLpro proceeds in a similar fashion to latent vector optimization. Differentiable scores are used to repeatedly evaluate the ligand's bound pose, and the derivatives of these scores with respect to the molecular coordinates are used to update the pose in the protein pocket with a gradient-based optimizer.

We use the L-BFGS optimizer in Pytorch~\cite{paszke2019pytorch} to perform coordinate optimization. Our optimization algorithm is implemented with Pytorch and is parallelizable on a GPU. Using structural refinement instead of re-docking each individually generated ligand using an auto-docking software such as AutoDock Vina~\cite{trott2010autodock} or QuickVina2~\cite{alhossary2015fast} affords us an increase in overall computational efficiency.In this work, we only use one combination of evaluators to perform coordinate optimization: torchvina and ANI2x~\cite{devereux2020extending}. 

To balance the validity of relaxed molecules with docking performance, We consider various combinations of intra- and inter-molecular forces derived from the torchvina and ANI2x potentials. We run structural refinement on a set of 200 ligands generated by IDOLpro when seeded with pockets in our validation set (see SI for details). We tabulate the results in~\cref{tab:structural-refine} and include metrics when QuickVina2~\citeS{alhossary2015fast} is used for re-docking. For each setting, we compute the the average Vina and top-10\% Vina scores of relaxed molecules, as well as percent of molecules which remain valid according to our validity checks after structural refinement is applied. We initially found that invalid molecules were often caused by bonded atoms being pulled apart during structural refinement. To remedy this, we add an $L_1$ penalty for violating bonds in the molecule. We include the effect of adding this penalty in \cref{tab:structural-refine}.

\subsection{Regressor Guidance}
Classifier guidance was first proposed in Ref.~\cite{dhariwal2021diffusion}. Classifier guidance can be straightforwardly re-interpreted for regressor guidance, which has been implemented for molecular generation in a number of works~\cite{weiss2023guided,lee2023exploring,ziv2024molsnapper}. To apply regressor guidance to molecular generation, a regressor is trained to predict the physicochemical score for the final generated molecule given an intermediate latent vector in the denoising process. In particular, given some molecule $\mathbf{z}$ produced by a generative diffusion model, and some physicochemical score $f: Z \to \mathbb{R}$, a regressor is trained to approximate $f(\mathbf{z})$ with $\hat{f}_{\theta}(\mathbf{z}_t, t)$.

To implement regressor guidance in DiffSBDD, we train a regressor to predict the SA score of a molecule produced by DiffSBDD given an intermediate noisy latent vector. To do so, we run DiffSBDD, and save the entire trajectory of latent vectors, $\mathbf{z}_{\ell}^T, \mathbf{z}_{\ell}^{T-1}, \hdots, \mathbf{z}_{\ell}^0$. We then train an equivariant graph neural network (EGNN)~\cite{satorras2021n} to minimize the $L_2$ loss between its prediction of the SA score given an intermediate noisy latent vector, and the SA score of the final molecule. More information about the training of the EGNN regressor can be found in the SI.

At each step in the denoising process, a gradient term from the trained regressor is added to the sampled latent vector, i.e.,
\begin{align*}
    \tilde{\mathbf{z}}^{\ell}_{t-1} &\gets q(\mathbf{z}_{t-1}^{\ell} | \mathbf{z}_{t}^{\ell}, \mathbf{z}^p) \\
    \mathbf{z}^{\ell}_{t-1} &\gets \tilde{\mathbf{z}}^{\ell}_{t-1} + \lambda \nabla \hat{f}_{\theta}(\mathbf{z}^{\ell}_{t}, t).
\end{align*}
The first equation samples a new ligand given the current noisy ligand and pocket atoms, and the second equation applies regressor guidance with a factor of $\lambda > 0$, maximizing the SA score of the molecule. In our experiment (\cref{subsec:reg-guide}) we set $\lambda = 0.5$.

\section{Data Availability}
Our source code is publicly available at \url{https://github.com/sandbox-quantum/idolpro}.
We used the following publicly available datasets for validation and testing:
CrossDocked2020
(\url{https://bits.csb.pitt.edu/files/crossdock2020/}), and
Binding MOAD (\url{http://www.bindingmoad.org/}).
Detailed testing splits for generating each of the test benchmarks used in this work are described in \url{https://github.com/pengxingang/Pocket2Mol/tree/main/data}. A subset of CrossDocked2020 was used to select hyperparameters as described in the SI, and is available at \url{https://github.com/mattragoza/LiGAN/tree/master/data/crossdock2020}.

\section{Author Contributions}
AK, KR, and TY conceived the study, designed IDOLpro, and planned the experiments. AK and KR implemented IDOLpro, ran experiments, and wrote the manuscript. EL implemented the lead optimization capability in IDOLpro and wrote the corresponding section in the manuscript. AR and TY advised and reviewed the manuscript.

\section{Conflicts of Interest}
There are no conflicts to declare.

\section{Acknowledgements}

The authors thank Andrea Bortolato, Andrew Wildman, Arman Zaribafiyan, Benjamin Shields, and Jordan Crivelli-Decker for their feedback on the manuscript.



\balance

\bibliography{refs} 
\bibliographystyle{rsc} 

\appendix
\setcounter{section}{18}
\setcounter{table}{0}
\setcounter{figure}{0}
\renewcommand{\thesection}{\Alph{section}}
\makeatletter
\renewcommand\thetable{S\@arabic\c@table}
\renewcommand \thefigure{S\@arabic\c@figure}
\makeatother
\section{Supplementary Information}

\subsection{Computational Resources}
All experiments in \cref{subsec:latent-opt}, \cref{subsec:compare-dl}, \cref{subsec:compare-ga}, \cref{subsec:lead-opt} were run on an AWS instance with 8 CPU cores and an NVIDIA A10G GPU with 24 GB of VRAM. In \cref{subsec:compare-vs} Virtual screening with QVINA on ZINC50K was run a compute-optimized AWS instance with 8 CPU cores, while IDOLpro was run on an AWS instance with 8 CPU cores and an NVIDIA A10G GPU with 24 GB of VRAM. Experiments in \cref{subsec:reg-guide} were run on a Google Cloud compute virtual machine instance with 8 cpu cores and an NVIDIA T4 GPU with 16 GB VRAM.

\subsection{More Details on Training the TorchSA Model}\label{si:torchsa-details}

\begin{table*}[ht!]
\centering
\begin{tabular}{c|c}
     PDB ID & Ligand ID \\ \hline
     2ah9 & cto \\
     5lvq & p2l \\
     5g3n & u8d \\
     1u0f & g6p \\
     4bnw & fxe \\
     4i91 & cpz \\
     2ati & ihu \\
     2hw1 & lj9 \\
     1bvr & geq \\
     1zyu & k2q \\
\end{tabular}
\caption{Validation set used to choose hyper-parameters in IDOLpro. All proteins from the test set of LiGAN~\cite{ragoza2022generating} were used, and a single protein pocket for each protein was selected at random.}
\label{si-tbl:val-set-hyperparam}
\end{table*}

\begin{table*}[ht!]
\centering
\begin{tabular}{c|c|c|c|c}
     Diffusion steps & Vina [kcal / mol] & SA & QED & Time [s] \\ \hline
     5 & $-6.25 \pm 2.08$ & $3.72 \pm 0.46$ & $0.53 \pm 0.10$ & $60.60 \pm 49.75$ \\
     50 & $-6.72 \pm 2.45$ & $3.92 \pm 0.58$ & $0.54 \pm 0.10$ & $196.96 \pm 102.97$  \\
\end{tabular}
\caption{Results when reducing the number of rollout steps from 50 to 5. The average Vina, SA, and QED across the validation set is reported.}
\label{si-tbl:acceleration}
\end{table*}

\begin{table*}[ht!]
\centering
\begin{tabular}{c|c|c|c}
     $t_{hz}$ & $\Delta$ Vina & $\Delta$ SA & $\Delta$ QED \\ \hline
     50 & -1.21 & -0.34 & -0.028 \\
     100 & -1.17 & -0.82 & 0.004 \\
     200 & -1.84 & -0.76 & 0.048 \\
\end{tabular}
\caption{Results when varying the optimization horizon $t_{hz}$ in IDOLpro. The difference in Vina, SA, and QED for the final optimized ligands produced by IDOLpro relative to the initial ligands produced by DiffSBDD are reported.}
\label{si-tbl:opt-horizon-tune}
\end{table*}

\begin{figure}[ht]
    \centering
    \includegraphics[width=\linewidth]{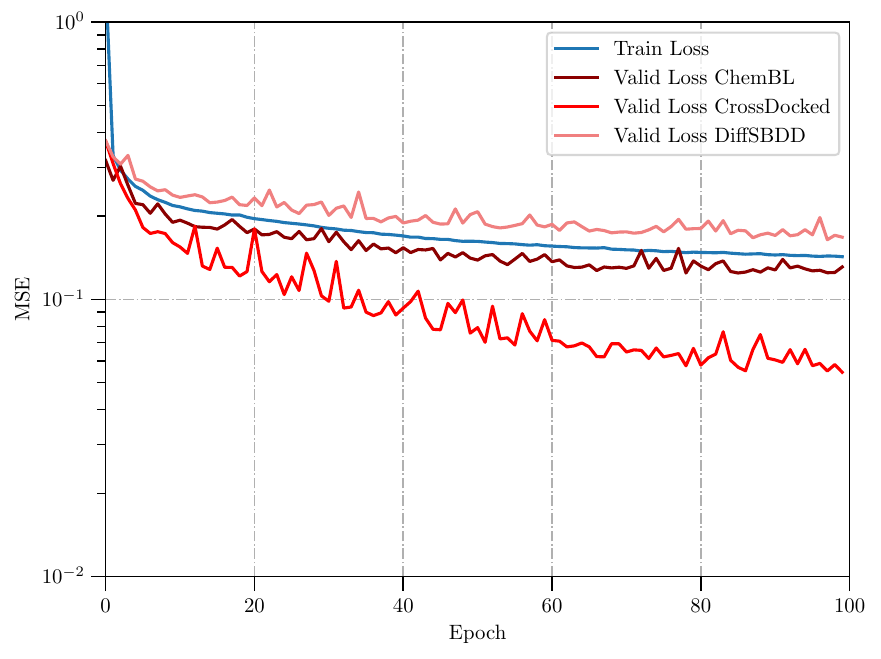}
    \caption{Training and validation curves for training PaiNN to predict the SA score on 3D atomic point clouds. The MSE at each epoch is plotted for the training set, the ChemBL validation set, the DiffSBDD validation set, and the CrossDocked validation set.} 
    \label{fig:torchsa_train}
\end{figure}

To train the SA model, we prepare a dataset consisting of all molecules with structural information in ChemBL~\citeS{zdrazil2024chembl} (2,409,270 structures), and ligands used to train DiffSBDD~\citeS{schneuing2022structure} on CrossDocked2020 ~\citeS{francoeur2020three} (183,468 structures). Although the SA score is fully determined by the chemical graph of a molecule, we keep molecules with different conformations from CrossDocked2020 to aid the model in learning the redundancy of pose in determining the SA score. To improve the model's performance on ligands produced by DiffSBDD, we generate nearly 1,000,000 (877,284) ligands with DiffSBDD which are included in the training data. We generate several ligands for each of the protein pockets in the DiffSBDD training set and then filter them using the same validity checks described in the Methods section. We put a higher emphasis on modelling ligands from CrossDocked2020 and DiffSBDD, sampling from one of these datasets during training with a $5\times$ higher likelihood than ChEMBL.

We train the polarizable atomic interaction neural network~\citeS{schutt2021equivariant} (PaiNN) from the Open Catalyst Project~\citeS{chanussot2021open,tran2023open} to predict the SA score given the atomic coordinates and atom types. To allow PaiNN to make predictions on atom types coming out of DiffSBDD, we encode atom types as one-hot vectors. We first optimize the hyperparameters of the model using Ray Tune~\citeS{liaw2018tune}. The hyperparameters chosen were \textit{num\_rbf} = 64 \textit{num\_layers} = 4, \textit{max\_neighbor} = 30, \textit{cutoff} = 8.0, \textit{hidden\_channels} = 256. We use a 95\%/5\% training/validation split for each dataset. The model is trained for 100 epochs to minimize the MSE loss with the AdamW optimizer~\citeS{loshchilov2017decoupled} with a learning rate of $5\times10^{-4}$. Training and validation curves are plotted in \cref{fig:torchsa_train}.

\subsection{More Details on Training the Regressor Guidance Model }\label{si:egnn-details}

To train the regressor used in \cref{subsec:reg-guide}, we prepare a dataset of latent vector trajectories from the reverse diffusion process of DiffSBDD. The training set for DiffSBDD consists of 100,000 protein-ligand pairs from the CrossDocked~\citeS{francoeur2020three} dataset. Each reverse diffusion trajctory of DiffSBDD consists of 500 denoising diffusion steps. In order to reduce the size of the training data, we take a random 10\% sample of DiffSBB's training data, resulting in 10,000 protein pockets which can be used to seed the generation of DiffSBDD. For each protein pocket, we generate a full reverse diffusion trajectory (500 denoising diffusion steps). Each latent vector in the trajectory $\mathbf{z}_{\ell}^t$ is labeled with the SA score of the final generated molecule, i.e., $SA(\mathbf{z_{\ell}^0})$. This results in 5,000,000 data points for training the regressor.

\begin{figure}[ht]
    \centering
    \includegraphics[width=\linewidth]{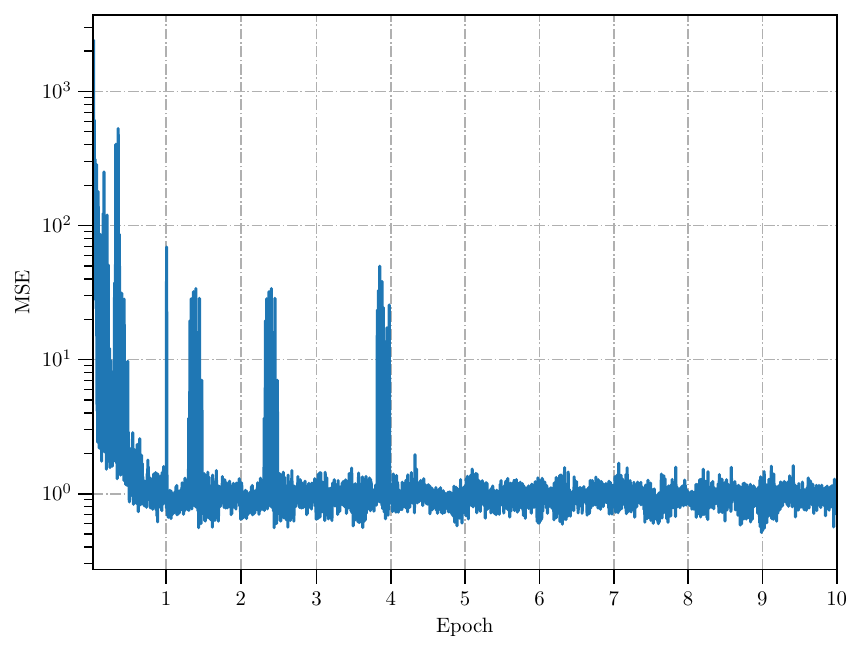}
    \caption{Loss curve for training EGNN to predict the SA score of molecules produced by DiffSBDD given intermediate noisy latent vectors. The MSE at each step is plotted.} 
    \label{fig:egnn_train}
\end{figure}

We train an equivariant graph neural network (EGNN)~\citeS{satorras2021n} with 4 layers and 256 hidden channels to predict the SA score of the final DiffSBDD molecule given an intermediate noisy latent vector. We also supply the model with the corresponding timestep in the reverse diffusion process. The model is trained for 10 epochs to minimize the MSE loss with the Adam optimizer~\citeS{loshchilov2017decoupled} with a learning rate of $1\times10^{-4}$. The training curve is plotted in \cref{fig:egnn_train}.

\subsection{Hyperparameter Tuning}
\label{subsec:hyperparam}

\subsubsection{Accelerating Diffusion}\label{si-subsubsec:accelerate}

In DiffSBDD, the models are trained to generate ligands over 500 reverse diffusion steps according to some noise schedule. At each time step an equivariant network takes in the noised coordinates and atom types, as well as the time step, and returns a denoised representation of the atoms and coordinates~\citeS{schneuing2022structure}. One can reduce the number of reverse diffusion steps by skipping time steps in the noise schedule.

In IDOLpro, the majority of the reverse diffusion process is run to generate $\mathbf{z}_{t_{hz}}$, i.e., to seed the initial latent vectors. When optimizing $\mathbf{z}_{t_{hz}}$, the rollout of $\mathbf{z}_{t_{hz}}, \hdots, \mathbf{z}_0$ needs to be repeated many times. We run an experiment to determine whether we can reduce the number of steps during this final rollout 10-fold without a significant degradation in performance. We run an experiment with the smallest horizon considered $t_{hz} = 50$. Results are shown in table~\cref{si-tbl:acceleration}. Running the rollout with reduced diffusion steps results in over a $3\times$ speedup, with a slight decrease in average Vina score (< 0.5 kcal /mol), while preserving average SA and QED. We adopt this setting in our pipeline for this reason.

\subsubsection{Tuning Optimization Horizon}\label{subsubsec:hyperparam-horizon}
We tune the value of the optimization horizon, ${t_{hz}}$, to optimize both the Vina and SA scores of generated molecules. We consider $t_{hz} \in {50, 100, 200}$. For each setting of the optimization horizon, we track the difference in Vina score, SA score, and QED. Results are reported in \cref{si-tbl:opt-horizon-tune}. Based on these results we set the optimization horizon to 200, since that setting resulted in by far the best difference in Vina score and QED, albeit a slightly worse improvement in SA score relative to setting $t_{hz}=100$.

\subsubsection{Structural Refinement with Torchvina and ANI2x}
\label{subsubsec:hyperparam-dock}
We use the following parameters in the L-BFGS optimization algorithm: \textit{max\_iter}=100, \textit{tolerance\_grad}=$10^{-3}$, and \textit{line\_search\_fn}=``strong\_wolfe". On top of using intra- and inter-molecular forces derived from the ANI2x~\citeS{devereux2020extending} and torchvina potentials, we inlcude an $L1$ penalty for violating the bonds in the molecule produced by IDOLpro. To do so, we use ASE's~\cite{larsen2017atomic} natural cutoffs. We find that setting a weight of 0.01 on this $L1$ penalty result in the best balance of Vina score and validity.

\subsubsection{Stopping Criteria, Backtracking, and Decaying Learning Rate}
\label{si-subsec:latent-opt-extra}
We use per-parameter options in Pytorch~\citeS{paszke2019pytorch} to allow for individualized learning rates for different ligands. For each ligand, we optimize it with Adam with the chosen hyperparameters. We optimize each latent vector for 10-200 optimization steps. Often, during latent vector optimization, a ligand will be pushed to a part of latent space such that it becomes invalid. In such a case, we attempt to generate a ligand 10 times with the given latent vector. If after 10 attempts, reverse diffusion has not produced a valid ligand, we backtrack to the previous latent vector in the optimization trajectory, reduce the learning rate by a factor of 10, and restart the optimization. If at another point in the optimization, with the reduced learning rate, another latent vector fails to generate a valid ligand over 10 attempts, the optimization of that trajectory is stopped.

\subsection{Visualization of Latent Vectors}
\begin{figure}[ht]
    \centering
    \includegraphics[width=\linewidth]{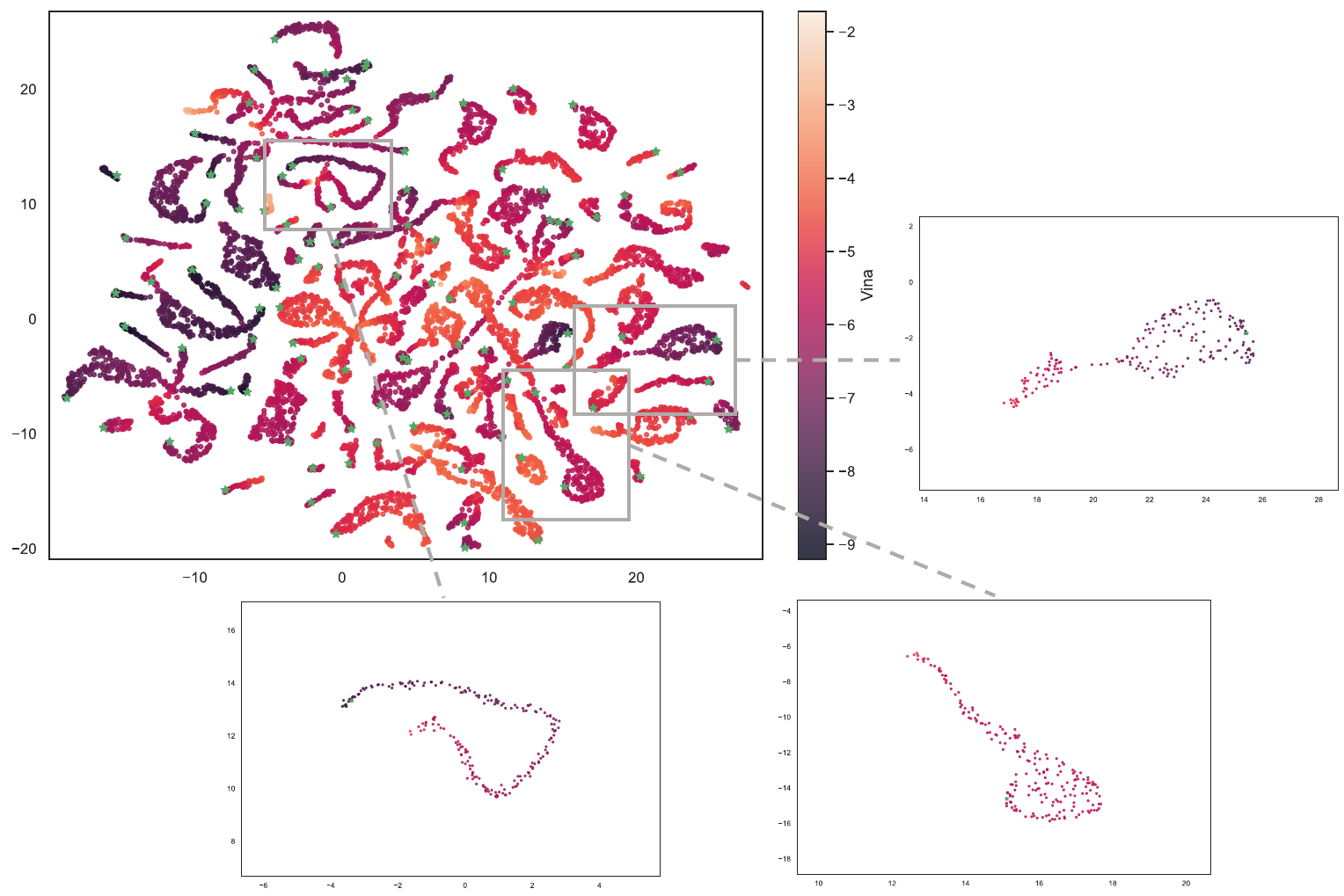}
    \caption{Latent vector visualizations of IDOLpro when generating ligands for 14gs. The points are coloured by Vina score (darker implies lower scores), and a green star marks the end of each optimization trajectory.}
    \label{fig:latent_viz}
\end{figure}

Latent vector visualization was performed with UMAP~\citeS{mcinnes2018umap} and visualized in~\cref{fig:latent_viz}.

\bibliographystyleS{rsc}
\bibliographyS{supp}

\end{document}